\documentclass[preprints,article,accept,moreauthors,pdftex]{mdpi} 

\usepackage{amsmath,amstext}
\usepackage[T1]{fontenc}
\usepackage{amssymb}
\input{epsf}
\usepackage{graphicx}
\usepackage{slashed}

\usepackage{hyperref}
\usepackage{amsmath}
\usepackage{amssymb}
\usepackage{mathtools}
\usepackage{bm}
\usepackage{cleveref}
\usepackage{tensor}
\usepackage{braket}
\usepackage{mhchem}
\usepackage{physics}
\usepackage{upgreek}

\usepackage{color}

\newcommand{\spc}{\quad \quad \quad}
\newcommand{\paral}{\mathbin{\!/\mkern-5mu/\!}}

\def\be{\begin{equation}}
\def\ee{\end{equation}}
\def\beq{\begin{eqnarray}}
\def\eeq{\end{eqnarray}}

\firstpage{1} 
\pubvolume{xx}
\issuenum{1}
\articlenumber{5}
\pubyear{2020}
\copyrightyear{2020}
\history{}

\continuouspages{yes}

\pdfoutput=1

\Title{Superfluid dynamics in neutron star crusts: the Iordanskii force and chemical gauge covariance} 

\Author{
Lorenzo Gavassino\orcidA{}, Marco Antonelli\orcidB{} and Brynmor Haskell \orcidC{}
}

\AuthorNames{Lorenzo Gavassino, Marco Antonelli, Brynmor Haskell}

\address[1]{%
Nicolaus Copernicus Astronomical Center of the Polish Academy of Sciences, Bartycka 18, 00-716 Warszawa, Poland 
}

\corres{Email: \url{lorenzo@camk.edu.pl} (L.G.), \url{mantonelli@camk.edu.pl}  (M.A.)}

\abstract{
We present a geometrical derivation of the relativistic dynamics of the superfluid inner crust of a neutron star. The resulting model is analogous to the Hall-Vinen-Bekarevich-Khalatnikov hydrodynamics for a single-component superfluid at finite temperature, but particular attention should be paid to the fact that some fraction of the neutrons is locked to the motion of the protons in nuclei. This gives rise to an ambiguity in the definition of the two currents (the normal and the superfluid one) on which the model is built, a problem that manifests itself as a chemical gauge freedom of the theory. To ensure chemical gauge covariance of the hydrodynamic model, the phenomenological equation of motion for a quantized vortex should contain an extra transverse force, that is the relativistic version of the Iordanskii force discussed in the context of superfluid Helium.  Hence, we extend the mutual friction model of Langlois et al. (1998) to account for the possible presence of this Iordanskii-like force. Furthermore, we propose that a better understanding of the (still not completely settled) controversy around  the presence of the Iordanskii force in superfluid Helium, as well as in neutron stars,  may be achieved by considering that the different incompatible results present in the literature pertain to two, opposite, dynamical regimes of the fluid system.
}


\begin{document}


\section{Introduction} 

The presence of superfluidity has a significant impact on the behaviour and evolution of neutron stars,
both on short timescales (where superfluidity leads to additional modes of oscillation, e.g. \cite{and_com_lang_2002,KhomenkoPhysRevD2019,chamel_modes_crust_2013}) and on secular timescales (where superfluidity affects the nuclear reactions responsible for  neutrino cooling~\cite{yako_cooling_rev}). 
Furthermore pulsar glitch models \citep{haskellmelatos2015} are typically based on extensions to the neutron star problem of the  two-fluid Hall-Vinen-Bekarevich-Khalatnikov (HVBK) hydrodynamics, usually employed to describe superfluid $^4$He \cite{donnelly_book,sonin_book_2016}.

A particularly important point in this hydrodynamic description concerns the so-called mutual friction, a dissipative force coupling the superfluid and normal parts of the system, which is directly related to the presence of quantised vortices, e.g.~\cite{langlois98,andersson_sidery_2006,GusakovHVBK,Andersson_Mutual_Friction_2016,Rau2020PhRvD,sourie_corepinn_2020MNRAS,antonelli2020MNRAS}. 
In addition to being the coupling mechanism giving rise to glitches, the mutual friction may also provide the main damping mechanism for various classes of neutron star oscillations and it is a likely candidate for limiting the growth of modes that are driven unstable by the emission of gravitational waves, see e.g. \cite{lind_mend_2000PhRvD,Haskell_rmodes_2009}.
Hence, it is important to develop hydrodynamic relativistic models of neutron star interiors that can consistently take into account for the multifluid nature of the system, and possibly include a consistent modeling of vortex mediated mutual friction. 

In particular, this work is devoted to extending the effective two-fluid description of neutron star hydrodynamics developed by \citet{langlois98}, focusing our attention on the inner crust, where a superfluid of delocalized neutrons (that will be present in the inner layers below the neutron drip threshold at $\sim 10^{11}\,$g/cm$^3$) coexists with the exotic ions comprising the crustal lattice and a neutralizing background of relativistic electrons \cite{chamel_review_crust}.

Clearly, such a two-fluid treatment does not account for a third independent current,
that of the electrons, which is of dominant importance for magnetic effects but makes a negligible contribution to the mass transport effects under consideration here. 
Since also the  stress anisotropy that can arise from the elastic solidity of the crust, or from strong magnetic fields, is neglected, this effective two-fluid description parallels the relativistic version of the HVBK  hydrodynamics described by \citet{GusakovHVBK}, valid for a single species superfluid at finite temperature. We also allow for a special feature of the inner crust layers, namely the clustering of some nucleons into ions, which gives rise to a fundamental ambiguity on the number of free neutrons, as well as on the number of confined nucleons that behave as an effectively normal fluid in the  inner crust. This ambiguity was discussed in depth by \citet{carter_macro_2006}, and takes the form of a ``chemical gauge freedom'' of the macroscopic hydrodynamic model, see also~\cite{Termo}.

Furthermore, we also aim to discuss in a fully relativistic setting the relationship between the dissipative behavior of the  macroscopic HVKB two-fluid model and the most general Phenomenological Equation for Vortex Motion (PEVM), which expresses the balance of all the lift and drag forces acting on a vortex. The PEVM provides a closure of the hydrodynamic system (see e.g. \cite{Galantucci2020EPJP} for the case of $^4$He) by defining  the form of the vortex mediated mutual friction \cite{andersson_sidery_2006,sourie_corepinn_2020MNRAS,antonelli2020MNRAS}.

Our decision of considering the most general form of PEVM is not a mere exercise. In fact, there is a long-lasting debate around the possible presence of the so-called Iordanskii force acting on quantized vortices in superfluid $^4$He, which is still not completely settled 
\cite{Wexler1997,Sonin_comment_1998,Volovik1998pc,geller2000,stone2000,Thouless2001,sourie2020IJMPB}. 
In the context of neutron star crusts this problem is, possibly, even more severe, as the physical meaning of the ``superfluid'' and ``normal'' components is not uniquely defined because of the ambiguity on the number of free neutrons mentioned above.
Therefore, we provide a discussion of the way such a possible Iordanskii force appearing in the PEVM depends on the gauge freedom involved in the choice of a chemical basis for the purpose of specifying which neutrons are considered to be free. 
We prove that the presence of a Iordanskii-type force is a formal necessity to ensure the covariance of the macroscopic hydrodynamic theory under chemical gauge.

The paper is organised as follows. 
 
In Sec \ref{sezione2} we introduce the two-fluid model of \citet{langlois98} for neutron-star matter; we analyse the model from a geometrical perspective and  derive, in this setting, the relativistic form for the vortex velocity, which turns out to be equivalent to  the one of \citet{GusakovHVBK}.
In Sec \ref{sezione3} we consider the most general PEVM  for a non-turbulent vortex configuration. 
This PEVM contains two force terms that were not included in the model of \citet{langlois98}, see also \cite{Andersson_Mutual_Friction_2016,GavassinoGeometria}: a  rescaled Iordanskii force and a drag force due to the superfluid component.

In Sec \ref{sezione4} we present the chemical gauge freedom described by \citet{carter_macro_2006}. 
We prove that, to respect the formal covariance established by this microscopic principle, the hydrodynamic theory must account for a generic Iordanskii force term of the form introduced in Sec \ref{sezione3}.
 
In Sec \ref{sezione5} we apply the analysis of \citet{Carter_Prix_Magnus} to the context of neutron stars: this allows us to derive the relativistic counterparts of both the Sonin-Stone \cite{sonin1997PhRvB,stone2000,sonin_book_2016} and the \citet{Thouless2001} results for the Jukowski lift force (i.e. the total transverse force on a vortex). We show how these two incompatible results arise from different assumptions on the value of the circulation of the relativistic normal momentum around a vortex.
 
Finally, in Sec \ref{sec6} we propose a relativistic extension of the thought experiment of \citet{Wexler1997}, that was designed to clarify which are the transverse forces on a vortex in  superfluid $^4$He (Wexler concluded that there is no transverse force proportional to the normal fluid velocity, i.e. that there is no Iordanskii force, see also \cite{geller2000}). 
This is in apparent conflict with the result of \citet{sonin1997PhRvB}  and  \citet{stone2000}, that predicted the presence of the Iordanskii force on a microscopic basis (i.e. scattering of phonons in the vicinity of a vortex, see also e.g. \cite{sonin_book_2016}). We propose that, in both $^4$He and neutron stars, this controversy on the presence of the Iordanskii force may be overcome by considering that the incompatible results of Thouless  and Sonin-Stone pertain to two different dynamical regimes.

Throughout the paper we adopt the spacetime signature $ ( - , +, + , + ) $, choose units with the speed of light $c=1$ and Newton's constant $G=1$, use greek letters $\nu$, $\rho$, $\sigma$... for coordinate tensor indexes and 
latin letters $a,b,c,...$ for tetrad tensor indexes.
  For anti-symmetrization we adopt the shorthand notation $A_{[\nu \rho]}= (A_{\nu \rho} - A_{\rho \nu})/2 \, $. 
The sign of the volume form is chosen according to the convention $\epsilon_{0123} = \sqrt{-g}$.
The Hodge dual of a generic p-form $f$ 
is~$ \, {\star} f_{\nu_1 ... \nu_{4-p} } = 
(1/p!) \, \epsilon^{ \lambda_1 ... \lambda_p }_{\phantom{\lambda_1 ... \lambda_p}\nu_1 ... \nu_{4-p}} 
\, f_{\lambda_1 ... \lambda_p} \,$.

\section{Relativistic  formulation of the HVBK hydrodynamics}
\label{sezione2}

We introduce the two-fluid model of \citet{langlois98} for the finite temperature hydrodynamic description of a neutron star, which is essentially a relativistic version of the Hall-Vinen-Bekarevich-Khalatnikov (HVBK) hydrodynamics \cite{GusakovHVBK}. We restrict our attention to the inner crust (although the model can be in principle applied also to the core).
The two-component model is an effective description, in the sense that the nucleons in the ions, the relativistic electrons and the thermal excitations are treated as a single normal species. 

Contributions to the stress-energy due to the possible presence of an elastic vortex-array will not be considered here \citep{baymChandler1983,Carter_KalbRamond1995,Andersson2020vortexElasticity}. 
Furthermore, despite the presence of a solid component, the model neglects the elastic stresses \citep{ander_elasticity_2019}, the  anisotropies and the  inhomogeneities due to the presence of the ions \citep{Carter2006IJMPD,Pethick2010PThPS,Kobyakov_pasta_2018}.  
However, since our focus is on the evolution of the superfluid component, the hydrodynamic equations for the effective normal component are of secondary importance, as they just serve to consistently close the system. Therefore, the fact that the normal component is treated as a fluid should not compromise the general validity of our results. 

Given the above set of simplifications, our formulation is formally equivalent to the relativistic extension of the HVBK hydrodynamics for a single-species superfluid at finite temperature derived by~\citet{GusakovHVBK}.

\subsection{The stress-energy tensor}

The model of \citet{langlois98} builds on three currents, $s^\nu$, $p^\nu$, $n^\nu$, which can be interpreted as the current of entropy, protons and neutrons, respectively (we will need to go back to the problem of assigning a microscopic interpretation to $n^\nu$ in subsection \ref{chemicalGaugeEEEEE}).
For the time being,  $n^\nu$ counts all the neutrons, including those in the ions, so that the total current associated with the conserved baryon number is 
\begin{equation}\label{barbie}
b^\nu = p^\nu +n^\nu \, .
\end{equation} 
The second law of thermodynamics and the baryon conservation are implemented as
\begin{equation}
\nabla_\nu s^\nu \geq 0  \spc  \nabla_\nu b^\nu =0 \, .
\end{equation} 
For simplicity, chemical transfusion ($\beta$ reactions) and heat conduction are neglected, namely
\begin{equation}\label{semplifico}
\nabla_\nu p^\nu =-\nabla_\nu n^\nu= 0  \spc   s^{[ \nu}p^{\rho ]}=0 \, .
\end{equation}
Following \cite{langlois98}, to construct the stress-energy tensor of the theory it is useful to assume that the two-fluid system can be described in terms of an appropriate Lagrangian function $\Lambda(s^\nu,p^\nu,n^\nu)$, see e.g. \cite{Termo} for the precise interpretation of $\Lambda$ as a thermodynamic potential. 
An  infinitesimal variation of $\Lambda$, resulting from a change of the components of the currents at fixed metric components, can always be written in the form 
\begin{equation}\label{deltaLal}
\delta \Lambda = \Theta_\nu \delta s^\nu + \chi_\nu \delta p^\nu +\mu_\nu \delta n^\nu \, ,
\end{equation}
so that the covectors $\Theta_\nu$, $\chi_\nu$ and $\mu_\nu$ are the conjugate momenta to $s^\nu$, $p^\nu$ and $n^\nu$, respectively. 
As discussed in \citet{Termo}, the second assumption in \eqref{semplifico} implies that there is a freedom in the choice of $\Lambda$, which allows us to introduce the 
normal four-velocity $u^\nu$ and the local temperature of the fluid $\Theta$ as
\begin{equation}
u^\nu := {s^\nu}/(-s_\rho s^\rho)^{1/2}  \spc  
\Theta :=-\Theta_\nu \, u^\nu \spc 
\Theta_\nu = \Theta \, u_\nu \, ,
\end{equation}
without altering the physical content of the theory. 
Finally, the stress-energy tensor $T\indices{^\nu _\rho}$ derived from the variations of $\Lambda$ reads
\begin{equation}\label{tinuro}
T\indices{^\nu _\rho} = \Psi \delta\indices{^\nu _\rho} + s\Theta  \, u^\nu u_\rho +p^\nu \chi_\rho + n^\nu \mu_\rho \, ,
\end{equation} 
where  the thermodynamic pressure, see e.g. \cite{Termo}, is
\begin{equation}
\Psi = \Lambda + s\Theta  -p^\nu \chi_\nu - n^\nu \mu_\nu  \, .
\end{equation}

\subsection{The superfluid vorticity}\label{sec:vorticity}
 
We now introduce another conserved (i.e. physical) quantity in the hydrodynamic model: the vorticity related to the superfluid component (in fact, the quantized vortices in the superfluid cannot decay). We start by noting that the energy-momentum conservation $\nabla_\nu T\indices{^\nu _\rho}=0$, together with the conditions in  \eqref{semplifico}, give 
\begin{equation}\label{secondolaw}
\Theta_\rho \nabla_\nu s^\nu + 2 s^\nu \nabla_{[\nu} \Theta_{\rho]}+2 p^\nu \nabla_{[\nu} \chi_{\rho]}+2 n^\nu \nabla_{[\nu} \mu_{\rho]} =0 \, .
\end{equation}
Now, if we define the hydrodynamic vorticity of the neutron fluid as (e.g. \cite{carter92,Carter_defects2000})
\begin{equation}\label{curluzzograsso}
\varpi_{\nu \rho} \,  := 2  \, \nabla_{[ \nu} \mu_{\rho ]} 
\, = \, \partial_\nu \mu_\rho - \partial_\rho \mu_\nu  \, ,
\end{equation}
we can contract \eqref{secondolaw} with $u^\rho$ and find that
\begin{equation}
\label{secondBoss}
\Theta \nabla_\nu s^\nu = \varpi_{\nu \rho} n^\nu u^\rho \, .
\end{equation}
This tells us that dissipation, for a fixed local value of the currents, is determined locally by the geometric form of the vorticity tensor $\varpi_{\nu \rho}$.

With the present formalism it is possible to describe superfluids either at the local irrotational level (i.e. on the so-called mesoscopic scale, see e.g. \cite{Carter_defects2000}), or on the smooth-averaged macroscopic scale by neglecting the (generally small) anisotropy induced by the quantized vortices. At the mesoscopic scale, the additional requirement that the neutrons are in a superfluid state is encoded into the model by imposing the covariant Josephson relation  \citep{anderson_considerations,carter_macro_2006,Termo}
\begin{equation}
\label{joseph}
\mu_\nu = \dfrac{k}{2\pi} \nabla_\nu \phi  \spc k=\pi \hbar \,  ,
\end{equation}
where $\phi$ is the gradient of the phase of the superfluid order parameter, which implies that $\mu_\nu$ is the relativistic version of Landau's superfluid velocity within a mass factor \cite{carter92}. 
The relation \eqref{joseph} is valid at the inter-vortex separation scale and, at this scale, it ensures the irrotationality constraint
\begin{equation}
\label{zerovort}
\varpi_{\nu \rho} =0 \, .
\end{equation}
From \eqref{secondBoss} we can conclude that the hydrodynamic model at the inter-vortex mesoscopic scale, far from vortex-core region, is non-dissipative.
Conversely, at a scale larger than the inter-vortex separation, the vorticity $\varpi_{\nu \rho}$ is interpreted as the result of an average over many vortices and $\varpi_{\nu \rho} \neq 0$. 

Following \citep{Carter_defects2000}, in the physical limit in which there is no turbulence, i.e. the vortices are locally parallel, the condition \eqref{joseph} implies that there are two orthonormal four-vectors $u_v^\sigma$ (timelike, future-oriented) and  $l^\sigma$ (spacelike), in symbols
\begin{equation}\label{condizio}
u_v^\sigma \, u_{v\sigma} = -l^\sigma l_\sigma =-1  \spc u_v^\sigma \, l_\sigma =0 \, ,  
\end{equation} 
such that the macroscopic vorticity of the fluid element can be expressed as
\begin{equation}\label{explixit}
\varpi_{\nu \rho} = k \mathfrak{N} \, \varepsilon_{\nu \rho \sigma \lambda} \, u_v^\sigma \, l^\lambda  \spc \text{and} \spc u^\sigma l_\sigma =0 \, .
\end{equation}
The non-negative scalar $\mathfrak{N}$ can be interpreted as the local density per unit area of vortex lines, as measured by an observer that is locally comoving with the lines. 
The vectors $l^\sigma$ and $u_v^\sigma$ are, respectively, the unit tangent vector to the vortex lines and the average four-velocity of the vortex lines as measured by a local observer moving with $u^\sigma$ (the average four-velocity of the entropy in the local fluid element).

Equation \eqref{explixit} shows that $\varpi_{\nu \rho}$ contains information about both the usual 3-vorticity and the instantaneous average velocity of the lines. 
To see this more explicitly, it is useful to introduce a right-handed tetrad $e_a = e_a^\nu \partial_\nu$, constructed in a way that $e_0=u$ and $e_3=l$. This tetrad defines the local Lorentz frame of an observer comoving with the entropy, whose third axis is locally aligned with the vortices. In this tetrad, we can decompose the velocity $u_v= u_v^a e_a$ as 
\begin{equation}
\label{uvintetrad}
u_v=\Gamma_v(e_0+\Delta^1 e_1 +\Delta^2 e_2) \, ,
\end{equation}
so that the components of the four-vorticity read
\begin{equation}
\label{terbufi}
\varpi_{ab} = k \mathfrak{N} \Gamma_v 
\begin{bmatrix}
   0 & \Delta^2 & -\Delta^1 & 0  \\
    -\Delta^2 & 0 & 1 & 0  \\
    \Delta^1 & -1  & 0 & 0  \\
    0 & 0 & 0 & 0
\end{bmatrix} \, .
\end{equation}
We see that fixing the value of $\varpi_{0j}$ (for a given 3-vorticity) is equivalent to fixing the velocity at which the vortices are moving.
Now, following~\cite{Termo}, we indicate the quantities measured in the entropy frame, which is defined by the four-velocity~$u^\nu$, with a~$T$ label. 
According to this notation, equation~\eqref{terbufi} also implies that the density of vortices in the entropy frame is 
\begin{equation}
\label{thermalDensityVortex}
\mathfrak{N}_{T} := \Gamma_v \mathfrak{N} \, .
\end{equation}
Thanks to the decomposition \eqref{explixit}, it is possible to  better understand the meaning of \eqref{secondBoss} and the implications of the second law. 
In fact, by using \eqref{explixit} in \eqref{secondBoss}, we obtain
\begin{equation}
\label{gringo}
\Theta \nabla_\nu s^\nu = k \mathfrak{N} \, \varepsilon_{\nu \rho \sigma \lambda} u_v^\sigma l^\lambda  n^\nu u^\rho \geq 0 \, .
\end{equation} 
This shows that there is no dissipation only if $u_v \in \text{span} \{ l, n,u \}$. 
Finally, rewriting \eqref{gringo} in the aforementioned tetrad, we obtain the condition
\begin{equation}
\label{bufaloscemo}
\varepsilon_{0jk3} \, u_v^j \, n^k \, \geq 0 \, .
\end{equation}
Hence, dissipation is possible if, in the frame of the entropy, there is a component of the  vortex 3-velocity which is orthogonal to the 3-velocity of the superfluid component. 

Finally, we remark that the mesoscopic model describing the fluid at the inter-vortex scale is non-dissipative, while in the macroscopic model, where the vorticity is averaged over a fluid element containing many vortices, a positive entropy production is possible. 
This implies that the processes leading to dissipation occur close to  the vortex cores, where the mesoscopic hydrodynamic description breaks down. Therefore, it is impossible to move directly from the mesoscopic to the macroscopic model without inserting some additional information about what happens near the core of the vortices, which is external to the mesoscopic model itself.

\subsection{Closure of the macroscopic model: the dynamics of vortices}

At the macroscopic scale, the model should completely define the evolution of 9 hydrodynamic degrees of freedom: the two independent currents, $n^\nu$ and $p^\nu$, and, given the collinearity condition \eqref{semplifico}, the scalar $s=\sqrt{-s^\rho s_\rho}$. 
Proton, neutron and energy-momentum conservation produce 6 equations of motion, so that 3 additional equations are needed. 
The information needed to close the system is a model for the average dynamics of the quantized vortices \cite{nemirovskii_closure_2020JLTP}, which can be specified by assuming a phenomenological relation of the type 
\begin{equation}
\label{laveramutual}
u_v^\rho  = u_v^\rho (s,p^\sigma,n^\sigma,l^\sigma,\mathfrak{N}_{T}).
\end{equation}
Equation \eqref{laveramutual} determines (for a given local state of the fluid) at which velocity the vortices in the fluid element move, as measured in the frame of the normal component (i.e. the entropy frame introduced in the previous subsection).

The relation \eqref{laveramutual} produces three (and only three) differential equations that provide a closure to the system. Too see this, let us insert \eqref{laveramutual} into the first equation in \eqref{explixit}, which gives a condition of the type
\begin{equation}
\varpi_{\nu \rho} = \varpi_{\nu \rho} (s,p^\sigma,n^\sigma,l^\sigma,\mathfrak{N}_{T}) \, .
\end{equation}  
These are 6 independent equations but the 3 space components $\varpi_{jk}$ are not dynamical equations, since no time derivative is involved. These 3 equations just need to be combined with the 2 conditions $u^\sigma l_\sigma=0$ and $l^\sigma l_\sigma=1$ to constrain uniquely the values of the 5 variables $l^\nu$ and $\mathfrak{N}_{T}$ in terms of the initial configuration of the fluid (essentially, they are the relativistic analogue of the Newtonian vorticity definition $\boldsymbol{\omega}=\nabla \times \boldsymbol{\mu}$).  
The 3 remaining equations for $\varpi_{0j}$ are dynamical: they can be rewritten as
\begin{equation}
\partial_t \mu_j = \partial_j \mu_t + \varpi_{tj} (s,p^\sigma,n^\sigma,l^\sigma,\mathfrak{N}_{T})
\end{equation}
and provide a closure to the system.
In fact, a delicate point in the construction of a model for macroscopic superfluid hydrodynamics is the derivation from microphysics of a prescription for \eqref{laveramutual}, which may be extracted from simulations of the average velocity of an ensemble of vortex lines (see e.g. \cite{Galantucci2020EPJP} for $^4$He case, or \cite{antonelli2020MNRAS} for a neutron star application). This should be done by solving a ``force balance equation'' for every vortex in a fluid element (i.e. by requiring that the forces acting on the single vortices vanish \cite{donnelly_book,andersson_sidery_2006,sonin_book_2016}), as will be further  discussed in Sec \ref{sezione3}.

\subsection{Geometric decomposition of the vortex velocity} 

The vortex velocity in equation \eqref{laveramutual} can be written in a form which is analogous to the one used  by \citet{GusakovHVBK} in the derivation of relativistic HVBK hydrodynamics.
Firstly, it is convenient to perform an orthogonal decomposition of the neutron current in the normal-frame,
\begin{equation}
\label{pizzobaldo}
n^\sigma = n^{T} u^\sigma + J^\sigma \, ,   \spc J^\rho u_\rho =0  \, , \spc n^{T} 
=-n^\rho u_\rho \, , 
\end{equation}
where $n^{T}$ is the neutron density  measured in the reference frame of the normal component. Now, let us assume that the three vectors $u$, $n$ and $l$ are linearly independent (note that if this were not the case, then from \eqref{gringo} we would conclude that there is no dissipation). 
Then, the four vectors $\{ u\, , \, J\, , \, l\, , \,  -{\star} (J \wedge u \wedge l)\}$ provide us with a convenient basis of the tangent space. In this basis, the vortex velocity $u_v$ can be expanded in components as
\begin{equation}
\label{velocitabella}
u_v^\nu = \Gamma_v \, u^\nu + u_v^{(J)} \, J^\nu +u_v^{(l)} \, l^\nu + \mathcal{D} \, \varepsilon^{\nu \rho \sigma \lambda} J_\rho u_\sigma l_\lambda \, ,
\end{equation}
where the Lorentz factor $\Gamma_v = -u_v^\sigma u_\sigma $ is the same as the one appearing in  \eqref{uvintetrad} because $e_0=u$. 
Contracting \eqref{velocitabella} with $l_\nu$ and recalling the defining relations \eqref{condizio} and \eqref{explixit}, we find the constraint
\begin{equation}\label{constr}
u_v^{(l)} = - u_v^{(J)} J^\nu l_\nu \, .
\end{equation}
We can also use this expression for the vortex velocity to see how the second law \eqref{gringo} looks like: inserting \eqref{velocitabella} into \eqref{gringo}, we obtain  
\begin{equation}
\Theta \nabla_\nu s^\nu = k \mathfrak{N} \mathcal{D} \, \varepsilon_{\nu \rho \sigma \lambda}  \, \varepsilon^{\sigma \mu \alpha \beta} J_\mu u_\alpha l_\beta l^\lambda  J^\nu u^\rho \geq 0 \, .
\end{equation}
This expression can be further  simplified (working in the tetrad we introduced in subsection \ref{sec:vorticity} the calculations become straightforward) to 
\begin{equation}
\Theta \nabla_\nu s^\nu = k \mathfrak{N} \mathcal{D} \big[ J^\nu J_\nu - (J^\nu l_\nu)^2 \big] \geq 0 \, .
\end{equation}
Note that, since both $J$ and $l$ are spacelike, the Cauchy-Schwartz inequality $J^\nu J_\nu \geq (J^\nu l_\nu)^2$ holds (recall that $l^\nu l_\nu=1$); therefore, the second law of thermodynamics leads to:
\begin{equation}
\label{sesso}
\mathcal{D} \geq 0 \, .
\end{equation}
Thus, we have verified that the dissipation of energy into heat manifests itself at the level of the vortex dynamics in a non-zero value of $\mathcal{D}$, in agreement with the analysis of Section \ref{sec:vorticity}. 

Now, let us go back to the general expression of $u_v$, equation \eqref{velocitabella}, and analyse the number of free parameters needed to completely specify the vortex velocity. 
Considering the constraint \eqref{constr}, the 3 coefficients $\Gamma_v,u_v^{(J)},\mathcal{D}$ are enough to fix $u_v^\sigma$ uniquely. However, these are not all independent, since the normalization condition $u_v^\nu u_{v\nu}=-1$ can be used to write $\Gamma_v$ in terms of the two remaining coefficients. Therefore, we have converted the problem of determining the law \eqref{laveramutual} into the search for a formula for the 2 coefficients $u_v^{(J)}$ and $\mathcal{D}$. 

Instead of using $u_v^{(J)}$ and $\mathcal{D}$, it is more convenient to express \eqref{velocitabella} by means of two rescaled kinetic coefficients $\slashed{\alpha}$ and $\slashed{\beta}$, that will be referred to as \textit{HVBK coefficients} and are defined as
\begin{equation}\label{pupazzi}
u_v^{(J)} = -\Gamma_v \slashed{\alpha} \spc  \mathcal{D}=\Gamma_v \slashed{\beta} \, .
\end{equation}
Thanks to the above definition, the expression for the vortex velocity (analogous to the one provided by  \citet{GusakovHVBK}) which closes the hydrodynamic system is 
\begin{equation}
\label{lacentrale}
u_v^\nu = \Gamma_v \big[ u^\nu -\slashed{\alpha} \, (J^\nu-J^\rho l_\rho l^\nu) + \slashed{\beta} \, \varepsilon^{\nu \rho \sigma \lambda} J_\rho u_\sigma l_\lambda \big] \, .
\end{equation}
Since \eqref{sesso} is equivalent to $\slashed{\beta} \geq 0$, we see that it is only the third term in the above equation that contributes to dissipation, while the term proportional to  $\slashed{\alpha}$ is dissipationless, see also the alternative discussion in \citep{GusakovHVBK}. 
We present the comparison with the relativistic HVBK formulation of \citet{GusakovHVBK} in Appendix \ref{AAA}, on the basis the equivalence established in \cite{Termo} between Carter's approach and the formalism of \citet{Son2001} and \citet{Gusakov2007}. 

\section{Phenomenological modelling of vortex dynamics}
\label{sezione3}

Microscopic  models for vortex dynamics do not provide directly an explicit form like \eqref{lacentrale} for $u_v$, which takes the form of a force balance equation in the surroundings of a vortex, e.g. \cite{andersson_sidery_2006,sourie_corepinn_2020MNRAS,sourie2020IJMPB,antonelli2020MNRAS}. 
The PEVM is thus typically modelled as an algebraic equation involving $u_v$ and the velocities of the two components (its most general form in the Newtonian limit is described, e.g., in Appendix A of \cite{antonelli2020MNRAS}, see also \cite{sourie2020IJMPB}). 

The general form of $u_v$ in \eqref{lacentrale} must be the solution to the PEVM: this allows to rewrite the coefficients $\slashed{\alpha}$ and $\slashed{\beta}$ in terms of the microscopic parameters in the PEVM, which are expected to be linked to physical processes in the vicinity of the vortex core \cite{barenghi1983JLTP,donnelly_book,sonin_book_2016}.

In this section we derive the most general form that the PEVM can have, demanding that it is consistent with the hydrodynamic formalism presented in the previous section. 
In addition, we show how to compute the coefficients $\slashed{\alpha}$ and $\slashed{\beta}$ directly from this generic relativistic form of the PEVM.

\subsection{Projection tensors and physical basis for vortex dynamics}

Before introducing the PEVM we need to define two tensors which will be of crucial importance in this section.
The plane spanned by $\{ u_v, l \}$ constitutes the kernel of the vorticity two-form: a generic vector $v \neq 0$ satisfies
\begin{equation}
\varpi_{\nu \rho} v^\nu =  k \mathfrak{N} \, \varepsilon_{\nu \rho \sigma \lambda} u_v^\sigma l^\lambda v^\nu=0
\end{equation}
if and only if it is a linear combination of $u_v$ and $l$. It can be shown that this plane coincides with the tangent space to the worldsheet drawn by the vortex line which crosses the spacetime point under consideration \cite{Carter_defects2000,GavassinoGeometria}.  

Now, we can use the $\{ u_v, l \}$ plane to define a unique orthogonal decomposition of a generic vector by introducing the two projectors
\begin{equation}\label{paralleEel}
{\paral \,}\indices{^\nu _\rho} = -u_v^\nu \, u_{v\rho} + l^\nu l_\rho 
\spc
{\perp}\indices{^\nu _\rho} = \delta\indices{^\nu_\rho} +u_v^\nu \,u_{v\rho} - l^\nu l_\rho
\spc
\delta\indices{^\nu _\rho} = {\paral \,}\indices{^\nu _\rho} + {\perp}\indices{^\nu _\rho}
\, .
\end{equation}
These two projection tensors can be written in terms of the macroscopic vorticity as~\citep{langlois98,GavassinoGeometria}
\begin{equation}\label{quadrico}
{\paral \,}\indices{^\nu _\rho} = \dfrac{\star \varpi^{\nu \lambda} {\star} \varpi_{\lambda \rho}}{(k\mathfrak{N})^2}  \spc {\perp}\indices{^\nu _\rho}  = \dfrac{ \varpi^{\nu \lambda}  \varpi_{ \rho  \lambda}}{(k\mathfrak{N})^2}  ,
\end{equation} 
where $\star$ denotes the Hodge duality operator. 
Equations \eqref{quadrico} immediately imply that any contraction of the kind $\varpi_{\nu \rho}\varpi^{\rho \lambda} \varpi_{\lambda \sigma}...$ must result in a tensor which is proportional either to $\varpi$ itself (if we contract an odd number of vorticity tensors) or to ${\perp}$ (if we contract an even number of vorticity tensors). 
The same holds with $\star \varpi$, which will always produce either a result proportional to $\star \varpi$ or to ${\paral}$.

In view of this fact, let us now take again an arbitrary vector $v$ and consider the four vectors
\begin{equation}\label{base}
\varpi^{\nu \rho}v_\rho 
\spc {\perp}\indices{^\nu _\rho} v^\rho 
\spc {\star} \varpi^{\nu \rho}v_\rho 
\spc {\paral \,}\indices{^\nu _\rho} v^\rho  \,  .
\end{equation}
If none of them is zero, they constitute an orthogonal basis of the tangent space which is adapted to the orthogonal decomposition defined in \eqref{paralleEel}. 
Since the contraction of any of these basis vectors with $\varpi$ (or $\star \varpi$) will always result in a vector which is proportional to another of these basis vectors (with a coefficient which does not depend on $v$), we can conclude that \eqref{base} is the only possible basis (up to numerical factors) which 
\begin{itemize}
\item is defined in a covariant way using only the 2-form $\varpi$ and the vector $v$, 
\item depends linearly on $v$,
\item is orthogonal for any choice of $v$.
\end{itemize}
Therefore, this basis is a local ``physical basis'' for vortex dynamics, so that each contribution to the general PEVM is expected to be proportional to one of the 4 expressions of \eqref{base}, constructed taking as $v$ any vector which is assumed to be relevant to the dynamics of a vortex. 

To better clarify how this works, we provide the following example. Consider the vectors~$u_v$ and~$l$, i.e. the vortex velocity and its direction, as measured by an observer moving with $u$. Since the only relevant vector involved in their definition is $u$, it is natural to decompose them on the basis \eqref{base} by taking $v=u$. In this way, one will find that $u_v$ and $l$ are proportional to respectively the ${\paral \,}\indices{^\nu _\rho} v^\rho$ and the ${\star} \varpi^{\nu \rho}v_\rho$ basis vector, confirming their interpretation as natural directions in the tangent space.
To show this, we contract \eqref{paralleEel} with $u^\rho$ and use the second equation in \eqref{explixit} to obtain
\begin{equation}\label{uabaobab}
u_v^\nu = \dfrac{{\paral \,}\indices{^\nu _\rho} u^\rho}{\Gamma_v} \, .
\end{equation}
If we contract this equation with $\star \varpi_{\sigma \nu}$ we find
\begin{equation}\label{bubbo}
{\star \varpi}_{\sigma \nu} u_v^\nu = \dfrac{\star \varpi_{\sigma \nu} \, {\paral \,}\indices{^\nu _\rho} u^\rho}{\Gamma_v} \, .
\end{equation}
However, since from \eqref{explixit} it immediately follows that
\begin{equation}
{\star \varpi}^{\sigma \nu} = -k\mathfrak{N}(u_v^\sigma l^\nu - u_v^\nu l^\sigma) \, ,
\end{equation}
it is straightforward to show that \eqref{bubbo} is equivalent to
\begin{equation}
l^\sigma = -\dfrac{\star \varpi^{\sigma \nu} u_\nu }{k \mathfrak{N} \Gamma_v} \, .
\end{equation}

\subsection{Formulation of the phenomenological equation of vortex motion (PEVM)}
\label{forrRza!}

In order to provide a microscopic interpretation to the coefficients $\slashed{\alpha}$ and $\slashed{\beta}$ appearing in the general expression for $u_v$, one needs to have a model for the vortex dynamics, the aforementioned PEVM. The most common formulation of the PEVM starts from the assumption that there is a balance of forces acting on the vortex lines (see e.g. the discussion in \cite{sourie2020IJMPB} for the $^4$He case), expressible as
\begin{equation}\label{forzanulla}
f_\nu =0 \, ,
\end{equation}
where $f_\nu$ is a total force per unit volume. To obtain a force per unit volume, one can start from the force per unit length acting on a single vortex line and then multiply it by the density of vortices per unit area \cite{andersson_sidery_2006,Andersson_Mutual_Friction_2016}. 
In a relativistic framework, it is useful to work with forces per unit volume because, being exchanges of four-momentum per unit spacetime volume (i.e. per unit volume and per unit time), they are four-vectors. 

In order to be as general as possible, we will not make, here, assumptions about $f_\nu$ based on microscopic arguments, but we will see how far we can go by using only principles of consistency of the theory: in this sense, the equation for vortex motion that we will find is purely ``phenomenological''.

Starting from the observation that the only two relevant vectors of the hydrodynamic model are $n$ and $u$, we apply the natural basis decomposition presented in the previous subsection and write $f$ as the sum of 8 contributions:
\begin{equation}\label{deCC}
f_\nu = \sum_{i=1}^8 f^{(i)}_\nu 
\end{equation}
with
\begin{equation}
\begin{split}
& f^{(1)}_\nu=z^{(1)} \, n^\rho \varpi_{ \rho \nu}  \spc  f^{(2)}_\nu =z^{(2)}  \, u^\rho \varpi_{ \rho \nu}   \\
& f^{(3)}_\nu=z^{(3)} \, {\perp}\indices{^\rho _\nu} n_\rho   \spc  f^{(4)}_\nu =z^{(4)} \,  {\perp}\indices{^\rho _\nu} u_\rho  \\
& f^{(5)}_\nu=z^{(5)}  \, n^\rho {\star}\varpi_{ \rho \nu}   \spc  f^{(6)}_\nu =z^{(6)} \,  u^\rho {\star}\varpi_{ \rho \nu} \\
& f^{(7)}_\nu= z^{(7)}  \, {\paral \,}\indices{^\rho _\nu} n_\rho  \spc  f^{(8)}_\nu =z^{(8)}  \,  {\paral \,}\indices{^\rho _\nu} u_\rho  \\
\end{split}
\label{stocazzone}
\end{equation}
The pre-factors $z^{(j)}$ are kinetic coefficients which encode the microphysics into the PEVM, namely equation \eqref{forzanulla}. 
We remark that the decomposition \eqref{deCC} is not an expansion of the total force $f$ on a basis (for which just 4 linearly independent vectors would be enough, e.g. $f^{(1)},f^{(3)},f^{(5)},f^{(7)}$) but it is a subdivision of the total force into the elementary parts that the most generic microscopic model is expected to produce. 
Therefore, contrarily to generic linear combination coefficients, the $z^{(j)}$ should not depend themselves on scalar products involving vectors of the form \eqref{base}. Instead, they should be real transport coefficients, namely kinetic coefficients which depend on the local thermodynamic properties of the fluid and on the vortex density. 

Now, if we contract equation \eqref{forzanulla} with the projectors introduced in the previous subsection, we obtain the system of equations
\begin{equation}\label{sistemoio}
   {\perp}\indices{^\nu _\rho} f_\nu = f^{(1)}_\rho + f^{(2)}_\rho +f^{(3)}_\rho+f^{(4)}_\rho =0 
  \spc  
   {\paral \,}\indices{^\nu _\rho} f_\nu = f^{(5)}_\rho + f^{(6)}_\rho +f^{(7)}_\rho +f^{(8)}_\rho =0 \, . 
\end{equation}
It is easy to show that it is possible to reduce the 8 equations \eqref{sistemoio} to a system of 5 independent equations. If we interpret them as hydrodynamic equations, recalling that we needed only 3 equations to completely fix the dynamics of the model, it is clear that \eqref{sistemoio} leads to an over-determination of the system (converting the 2 exceeding equations of motion into nonphysical constraints on the initial conditions). The only way to avoid this undesired outcome is to impose that either ${\perp}\indices{^\nu _\rho} f_\nu =0$ or ${\paral \,}\indices{^\nu _\rho} f_\nu =0$ is an identity, by setting all the relative $z^{(j)}$ to zero. 

In relativity, four-forces acting on point particles are orthogonal to the particles' worldlines. It is reasonable to expect that an analogous principle of orthogonality (this time to the worldsheet) applies also to the case of one-dimensional objects. This suggests us that the best candidate for being an identity is the relation ${\paral \,}\indices{^\nu _\rho} f_\nu =0$. It is also interesting to note that the terms $f^{(5)}$ and $f^{(6)}$ violate parity, thus, if their coefficients were not set to zero, they would give rise to very exotic dynamics with no possible analogue in laboratory superfluids. We therefore impose
\begin{equation}
z^{(5)}=z^{(6)}=z^{(7)}=z^{(8)}=0 \, .
\end{equation}
From this, we finally conclude that the generic PEVM should assume the form
\begin{equation}
\label{generaleBuf}
z^{(1)} \, n^\rho \varpi_{ \rho \nu} + z^{(2)}  \, u^\rho \varpi_{ \rho \nu} =  -z^{(3)} \, {\perp}\indices{^\rho _\nu} n_\rho - z^{(4)} \,  {\perp}\indices{^\rho _\nu} u_\rho \, .
\end{equation}

\subsection{Interpretation of the force terms in the PEVM}

Equation \eqref{generaleBuf} can be recast into a more physically transparent form. Let us assume that $z^{(1)} \neq 0$, so that we can rescale all the coefficients in such a way as to set
\begin{equation}
z^{(1)}=1 \, .
\end{equation}
Furthermore, we define three dimensionless coefficients $\mathcal{Z}$, $\mathcal{R}'$ and $\mathcal{R}$ via the relations
\begin{equation}
z^{(2)}=p \mathcal{Z} \spc  \dfrac{z^{(3)}}{k\mathfrak{N}}=-\mathcal{R}'  \spc \dfrac{z^{(4)}}{k\mathfrak{N}}=-(n^T+\mathcal{Z}p)\mathcal{R} \, ,
\end{equation} 
where $p:= \sqrt{-p^\nu p_\nu}$. With these definitions, the PEVM \eqref{generaleBuf} becomes
\begin{equation}
\label{persempresarai}
 n^\rho \varpi_{ \rho \nu} + \mathcal{Z}  \, p^\rho \varpi_{ \rho \nu} =  k\mathfrak{N}\mathcal{R}' \, {\perp}\indices{^\rho _\nu} n_\rho +k\mathfrak{N}(n^T+\mathcal{Z}p)\mathcal{R} \,  {\perp}\indices{^\rho _\nu} u_\rho \, .
\end{equation}
This expression contains the two terms $f^{(1)}$ and $f^{(4)}$ considered by \citet{langlois98}, which are usually taken into account in relativistic models of mutual friction for neutron star interiors, see e.g. \cite{Andersson_Mutual_Friction_2016,GavassinoGeometria}. 
The $f^{(1)}$ term, namely $ n^\rho \varpi_{ \rho \nu}$, has always been interpreted as the relativistic analogue of the Magnus force \citep{Carter_Prix_Magnus,langlois98,Andersson_Mutual_Friction_2016,GavassinoGeometria} and we will call it for definiteness \textit{Carter-Magnus force}.
The contribution $f^{(4)}$, i.e. the term proportional to $\mathcal{R}$, is the relativistic extension of the usual\footnote{
    In fact, a term of the form $f^{(4)}$, or its Newtonian analogue, is always included in all models for the dissipative PEVM in $^4$He, as well as in  neutron star interiors, see e.g. \cite{langlois98,andersson_sidery_2006,haskellsedrakian2017}. 
    Regarding neutron star modelling, in the special case $\mathcal{Z}=0$, or $z^{(2)}=0$, the definition of $\mathcal{R}$ reduces to the one of, e.g., 
    \cite{andersson_sidery_2006,Andersson_Mutual_Friction_2016,GavassinoGeometria,antonelli2020MNRAS}, 
    apart from a Lorentz factor which is negligible under the assumption of slow rotation.
}
drag force imparted on a vortex by the normal component.

However,  the generic PEVM \eqref{persempresarai} includes also the two additional forces $f^{(2)}$ and $f^{(3)}$, which are typically neglected, e.g.  \cite{langlois98,Andersson_Mutual_Friction_2016,GavassinoGeometria}.
The force $f^{(2)}$, namely the term proportional to $\mathcal{Z}$, has the same structure as the Carter-Magnus force but, instead of the neutron current, there is the normal current. 
This is the structure of a Iordanskii-type force\footnote{
As distinct from the Magnus force, the  Iordanskii force arises when a quantum vortex moves with respect to the heat bath made of quasiparticle excitations.  
\citet{Volovik1998pc} first proposed that  the interaction of quasiparticles with the velocity field around a vortex resembles the interaction of matter with the gravitational field induced by  a spinning cosmic string (gravitational Aharonov-Bohm effect). This analogy  allowed \citet{stone2000} to calculate the asymmetry in the scattering of phonons on the vortex and the associated Iordanskii transverse force, in agreement with the previous work of Sonin, see e.g.\cite{sonin1997PhRvB}. 
These results, in principle, would allow to set the value of $z^{(2)}$.
Our Generalized Iordanskii force, however, also includes the possibility of having additional transverse contributions  (e.g. a sort of ``transverse drag'', possibly due to quasiparticles other than phonons \cite{barenghi1983JLTP,donnelly_book}). For this reason, and because of the chemical gauge-covariance issue to be discussed in section \ref{ZZZZZaZZZu},  the coefficient $z^{(2)}$ and its temperature dependence are not specified. 
}, 
with a prefactor $\mathcal{Z}$ which can be set freely, so that we will refer to it as a \textit{Generalised Iordanskii force}. 
The force $f^{(3)}$, i.e. the one proportional to $\mathcal{R}'$, is a possible drag imparted by the neutron component (see, e.g., appendix A of \cite{antonelli2020MNRAS}).

If we contract \eqref{persempresarai} with $u^\nu$ and employ \eqref{secondBoss}, we obtain
\begin{equation}\label{ssssSsecondddLav}
\Theta \nabla_\nu s^\nu = k\mathfrak{N} \bigg[ \mathcal{R}' {\perp}_{\nu \rho} u^\nu n^\rho +(n^T+\mathcal{Z}p) \mathcal{R} {\perp}_{\nu \rho} u^\nu u^\rho \bigg] \geq 0 \, .
\end{equation}
Therefore, assuming that $n^T+\mathcal{Z}p>0$, the simplest way of ensuring the validity of the second law is to impose that
\begin{equation}\label{simplestttt}
\mathcal{R}'=0  \spc  \mathcal{R} \geq 0 \, .
\end{equation}
The validity of \eqref{simplestttt}, which implies the absence of the drag force $f^{(3)}$ exerted on the vortex by the superfluid current, is usually assumed in almost all models of mutual friction in neutron star interiors \citep{langlois98,chamel_review_crust,haskellsedrakian2017}, see also the discussion in \citep{antonelli2020MNRAS}.
However, since at this level there is no formal reason to rule out the case $\mathcal{R}' \neq 0$, for the sake of completeness we will retain the $f^{(3)}$ term also in the next subsection.

\subsection{Computation of the HVBK coefficients from the PEVM}
\label{HVBKcacloili}

Now that we have the general form of the PEVM, namely \eqref{persempresarai}, we can use it to compute the HVBK coefficients $\slashed{\alpha}$ and $\slashed{\beta}$ in terms of the transport coefficients $\mathcal{Z}$, $\mathcal{R}'$ and $\mathcal{R}$. The case $\mathcal{Z}=\mathcal{R}'=0$ has already been analysed in \cite{Andersson_Mutual_Friction_2016} and \cite{GavassinoGeometria}. 
We extend this result by including also the possible contribution of the Generalised Iordanskii force (whose meaning will be analysed in detail in the next sections) and the  drag with the neutron current.

To determine the HVBK coefficients, we work in the tetrad  introduced in subsection \ref{sec:vorticity}, with the further condition that 
\begin{equation}
e_2 = \dfrac{J -g(J,l) \, l}{\sqrt{g(J,J)-g(J,l)^2}}. 
\end{equation}
In App \ref{tettru} we summarize the properties of this tetrad and give the explicit components' expressions for the relevant vectors of the model in this physical basis. Now, let us define the \textit{effective neutron density} 
\begin{equation}
\label{neFFF}
n^T_{\text{eff}}:= n^T+\mathcal{Z}p \, ,
\end{equation}
that allows us to absorb the Generalised Iordanskii force into an effective total Carter-Magnus force. In fact, we can rewrite equation \eqref{persempresarai} in the tetrad as 
\begin{equation}
\label{pufficosa}
(n^T_{\text{eff}} u^a +J^a)\varpi_{ab} = 
k \mathfrak{N} {\perp}_{ab}(\mathcal{R}' n^a + \mathcal{R} n^T_{\text{eff}} u^a) \, .
\end{equation}
Under the assumption of non-relativistic relative speeds, the above equation written in components leads to (see App \ref{tetraddizzareIldrag} for the proof)
\begin{equation}\label{bellatutto}
\begin{split}
&  n^T_{\text{eff}} \, \slashed{\alpha} +1= \mathcal{R}_{\text{eff}} n^T_{\text{eff}} \, \slashed{\beta}  \\
&  -n^T_{\text{eff}} \, \slashed{\beta}= \mathcal{R}' + \mathcal{R}_{\text{eff}} n^T_{\text{eff}}
\, \slashed{\alpha} \,  ,
\end{split}
\end{equation}
where we have introduced 
\begin{equation}
\label{Reff}
\mathcal{R}_{\text{eff}} = \mathcal{R} + \dfrac{n^T}{n^T_{\text{eff}}} \mathcal{R}' \, .
\end{equation}
After solving this system, we finally obtain the explicit expression for $\slashed{\alpha}$ and $\slashed{\beta}$,
\begin{equation}
\label{alpha_beta}
 \slashed{\alpha} =-\dfrac{1}{n^T_{\text{eff}}} \, \dfrac{1  + \mathcal{R}_{\text{eff}} \mathcal{R}'}{1+ \mathcal{R}_{\text{eff}}^2} 
\spc  \slashed{\beta} = \dfrac{1}{n^T_{\text{eff}}} \dfrac{ \mathcal{R}_{\text{eff}} - \mathcal{R}'}{1+ \mathcal{R}_{\text{eff}}^2}. 
\end{equation}
The second law, that is  equivalent to $\slashed{\beta} \geq 0$, is respected not only if \eqref{simplestttt} is valid, but also whenever
\begin{equation}
\mathcal{R}_{\text{eff}} -\mathcal{R}' \geq 0 \, ,
\end{equation}
provided that $n^T_{\text{eff}}>0$.
Now, let us specialize our analysis to the case $\mathcal{R}'=0$, i.e. no drag with the superfluid component: the HVBK coefficients $\slashed{\alpha}$ and $\slashed{\beta}$ boil down to the usual result (see, e.g.,~\cite{andersson_sidery_2006,chamel_review_crust,haskellsedrakian2017,GusakovHVBK,GavassinoGeometria,antonelli2020MNRAS})
\begin{equation}
\slashed{\alpha} =-\dfrac{1}{n^T_{\text{eff}}} \, \dfrac{1 }{1+ \mathcal{R}^2}  \spc \slashed{\beta} = \dfrac{1}{n^T_{\text{eff}}} \dfrac{ \mathcal{R}}{1+ \mathcal{R}^2} \, .
\end{equation}
This result also tells us that the presence of the Iordanskii-type term plays only the role of replacing $n^T$ with the effective neutron density $n^T_{\text{eff}}\,$.  The main goal of the next section is to understand the formal origin of this mechanism.

Finally, we explicitly write the HVBK-like friction force acting on the superfluid component of the model, which depends on the HVBK coefficients $\slashed{\alpha}$ and $\slashed{\beta}$.
Recalling \eqref{lacentrale} and \eqref{curluzzograsso}, we contract equation \eqref{explixit} with $u^\nu$ to obtain (for the case $\mathcal{R}'=0$)
\begin{equation}\label{pevminverittamutua}
    u^\nu (\partial_\nu \mu_\rho - \partial_\rho \mu_\nu) = \dfrac{k \mathfrak{N}_T}{n_{\text{eff}}^T (1+\mathcal{R}^2)} \bigg[ u^\nu \varepsilon_{\nu \rho \sigma \lambda} J^\sigma l^\lambda - \mathcal{R}(J_\rho - J_\lambda l^\lambda l_\rho) \bigg],
\end{equation}
which is presented in the same form as equation (56) of \citet{GusakovHVBK}. 
The above expression is a convenient way of recasting the PEVM into a form that clearly identifies the hydrodynamic force (the right-hand side) acting on the superfluid component. In fact, equation \eqref{pevminverittamutua} is directly formulated as a system of 3 independent first-order differential equations for the momentum, which can be used to describe the momentum transfer in relativistic models for pulsar-glitches and neutron star oscillations. Furthermore, in the next section we will prove that this expression is also invariant under a change of chemical basis.

\section{The problem of the chemical basis}
\label{sezione4}

We have shown that a generic PEVM can, in principle, contain both a Carter-Magnus force and a Generalised Iordanskii force. However, \citet{langlois98} implicitly assumed that the Generalised Iordanskii force term should necessarily vanish, see also \cite{Andersson_Mutual_Friction_2016}. 
This assumption was suggested by the straightforward application of the action principle from which the hydrodynamic model arises and by the requirement that the macroscopic mutual friction should be in close relation with the forces acting on a single vortex (i.e. the forces of the PEVM). 

In this section we review the standard argument for the absence of the Generalised Iordanskii force. 
However, we will find that such a force is in fact necessary to ensure the chemical gauge-covariance of the macroscopic theory.

\subsection{The argument for the absence of the Generalised Iordanskii force}
\label{eulero}

\citet{langlois98} have shown that, if we apply the action principle of \citet{Carter_Starting_Point} by taking $p^\nu$ and $n^\nu$ as free currents, then, in addition to the energy-momentum conservation  \eqref{secondolaw}, we also obtain the Euler-Lagrange equation for the neutrons (see also \cite{Termo})
\begin{equation}\label{pulitoeuler}
n^\rho \varpi_{\rho \nu} =0 \, .
\end{equation}
This action principle, however, produces only a non-dissipative hydrodynamic model and is not sufficient to specify all the equations in the dissipative regime \cite{Termo}. 
To include dissipation, then, equation \eqref{pulitoeuler} should be replaced by
\begin{equation}
\label{batuffolo}
n^\rho \varpi_{\rho \nu} = f^n_\nu \, ,
\end{equation} 
where $f^n_\nu$ is the macroscopic  mutual friction. 
In the absence of turbulence, it is natural to identify the left-hand side with the Magnus force per unit volume (i.e. the Magnus force on a single vortex times the vortex density) and the right-hand side with some drag force  per unit volume, see e.g. \cite{andersson_sidery_2006,sourie2020IJMPB,antonelli2020MNRAS,nemirovskii_closure_2020JLTP}. 


Hence, by considering that  at the  mesoscopic scale the drag force is assumed to have the form $f^{(4)}$, we are lead to postulate a PEVM of the form \eqref{batuffolo}  but with $f^n_\nu =-f^{(4)}_\nu$, namely
\begin{equation}
\label{nuvarpi}
n^\rho \varpi_{\rho \nu} = k\mathfrak{N}n^T\mathcal{R} \,  {\perp}\indices{^\rho _\nu} u_\rho \, .
\end{equation}
We see that this final equation does not contain a Generalised Iordanskii force acting on vortices.
This happens because the action principle naturally leads to identify \eqref{pulitoeuler} with the non-dissipative  limit of the hydrodynamic model. In the following sections we will explore the shortcomings of this postulate, but we  first need to introduce the problem of the chemical gauge.


\subsection{The chemical gauge} 
\label{chemicalGaugeEEEEE}

We define a change of chemical basis of the kind introduced by \citet{Carter_Starting_Point}.
In general, it is always possible to introduce two new currents $\tilde{p}^\nu $ and $\tilde{n}^\nu$ as \begin{equation}
\label{tildiamo}
  p^\nu = (1-a)\tilde{p}^\nu  \spc  n^\nu = \tilde{n}^\nu + a \tilde{p}^\nu 
  \spc  b^\nu = p^\nu+n^\nu  = \tilde{p}^\nu + \tilde{n}^\nu 
  \,   ,
\end{equation}
where $a \neq 1$ is an arbitrary constant. This transformation does not affect the physical current $b^\nu$,
meaning that \eqref{tildiamo} is just a formal redistribution of the baryons between the two species $p$ and $n$. Moreover, since $a$ is constant, these new currents are both conserved,
\begin{equation}
\nabla_\nu \tilde{p}^\nu = \nabla_\nu \tilde{n}^\nu =0 \, ,
\end{equation}
and the fact that $p^\nu$ is collinear to $\tilde{p}^\nu$ gives
\begin{equation}
s^{[\nu} \tilde{p}^{\rho]}=0 \, .
\end{equation}
Therefore,  $\tilde{p}^\nu$ and $\tilde{n}^\nu$ satisfy equations \eqref{semplifico}, exactly as the original $p^\nu$ and $n^\nu$, and their conjugate momenta are\footnote{
Consider the effect of the transformation \eqref{tildiamo} on the variation of $\Lambda$ given in  \eqref{deltaLal}.
} 
\begin{equation}
\label{tuttiinbarca}
\tilde{\chi}_\nu = (1-a)\chi_\nu + a \mu_\nu  \spc  \tilde{\mu}_\nu = \mu_\nu
\, .
\end{equation}
The second equation of \eqref{tuttiinbarca}, combined with the Josephson relation \eqref{joseph}, implies that at the mesoscopic level $\tilde{n}^\nu$ still defines a superfluid species. 
The quantity $\tilde{n}^\nu$ thus has all the original properties of $n^\nu$.

Finally, the energy-momentum tensor \eqref{tinuro} defined via the variations of $\Lambda$ is not affected by this redefinition of the currents in \eqref{tildiamo}, namely
\begin{equation}
\tilde{T}\indices{^\nu _\rho} = T\indices{^\nu _\rho} \, .
\end{equation} 
This can be  checked by verifying explicitly that
\begin{equation}
\tilde{p}^\nu \tilde{\chi}_\rho +\tilde{n}^\nu \tilde{\mu}_\rho = p^\nu \chi_\rho + n^\nu \mu_\rho \, .
\end{equation} 
Therefore, we have shown that if we replace $p^\nu$ and $n^\nu$ with $\tilde{p}^\nu$ and $\tilde{n}^\nu$ all the hydrodynamic equations of section \ref{sezione2} remain the same (we just need to place a tilde where needed). Moreover, the quantities $b^\nu$, $s^\nu$, $T\indices{^\nu _\rho}$, $\mu_\nu$, are  invariant under the chemical basis transformation (i.e. are not ambiguously defined, as they constitute the real physical content of the theory). 
In other words, the change of variables
\begin{equation}
\label{chemicalGauge}
(p^\nu,n^\nu)  \longrightarrow (\tilde{p}^\nu,\tilde{n}^\nu)
\end{equation}
can be considered as a sort of ``gauge transformation'' of the model which leaves the form of the equations of the theory unchanged and does not affect the physical quantities of the theory (namely 
$b^\nu$, $s^\nu$, $T\indices{^\nu _\rho}$, $\mu_\nu$, which are all related to physical conservation laws, as discussed in~\citet{Termo}).  

In the context of neutron star crusts, this ambiguity in the definition of the currents of the theory  has been discussed for the first time by \citet{carter_macro_2006}, who called the transformation \eqref{chemicalGauge} \textit{chemical gauge}. They also point out that this freedom of choosing the variables in the hydrodynamic model reflects a fundamental ambiguity that has a microscopic origin, which we summarize below. 

It is theoretically well-established that in the inner crust the protons are mostly confined in neutron-rich nuclei, or  nuclear pasta clusters, which are immersed into a fluid of dripped neutrons \cite{chamel_review_crust}. 
At the beginning of section \ref{sezione2}, we defined $n^\nu$ as the current which counts all the neutrons. However, below a certain energy threshold, the neutrons in the ions are not able to cross the potential barriers between the nuclei. This means that, on a sufficiently short time-scale, they behave as if they were effectively ``confined'' in the nuclei. 
Therefore, it should be possible to work with a conserved current $\tilde{p}^\nu$ of confined baryons, which includes also the neutrons that cannot escape the nuclear clusters, and with a current of free neutrons $\tilde{n}^\nu$. These two new currents can be obtained from equations \eqref{tildiamo}, by setting $a=(A-Z)/A$, with $A$ and $Z$ being the mass and the charge number of the ions. 
The number $A$, however, is not uniquely determined, as it depends on the energy threshold at which a neutron can be considered confined. 
In fact, there can be marginally bound states with intermediate penetration time scales that are macroscopically long but cosmologically short: this gives rise to a considerable ambiguity in the definition of the currents. 

A concrete example of gauge fixing is the choice of ``paired gauge'', $A=2Z$, which classifies as confined only the neutrons in the tightly bound states that are directly paired with corresponding proton states in the nuclei \cite{carter_macro_2006}. 
Clearly, there are several other possibilities and  analogous confinement problems are also expected in the pasta phases, although proper nuclei do not exist anymore.

In a more general perspective, which goes beyond the case of neutron-star crusts under consideration, we can interpret this ambiguity in the currents as a consequence of the fact that, since fermionic superfluidity is just the manifestation of the existence of an order parameter (the pairing gap) describing a correlation between fermions near the Fermi surface, there is not a physical separation of the particles into ``normal'' and ``superfluid'' ones. 
Hence, whether we choose to classify low energy particles (located deep inside the Fermi sphere), trapped in bound states with normal particles, as belonging to the superfluid species (as they have the same chemical composition) or to the normal species (as they are dynamically forced to behave as normal) is just a matter of taste. 

\subsection{The Generalised Iordanskii force is necessary to guarantee chemical gauge covariance}\label{ZZZZZaZZZu}

Let us go back to the argument for the absence of the Generalised Iordanskii force  presented in Sec \ref{eulero}. As we said, a straightforward application of the action principle using $p^\nu$ and $n^\nu$ as fundamental currents may lead to assume the non-dissipative equation of motion \eqref{pulitoeuler}. If we make the change of variables \eqref{tildiamo}, equation \eqref{pulitoeuler} becomes
\begin{equation}\label{alala}
( \tilde{n}^\rho + a \tilde{p}^\rho ) \tilde{\varpi}_{\rho \nu}=0 \, ,
\end{equation}
where we have used the relation $\tilde{\varpi}_{\rho \nu } = \varpi_{\rho \nu }$,
which is a direct consequence of \eqref{tuttiinbarca}.
On the other hand, the action principle gives 
\begin{equation}\label{ccmcmc}
\tilde{n}^\rho \tilde{\varpi}_{\rho \nu }=0  
\end{equation}
when the currents  $\tilde{p}^\nu$ and $\tilde{n}^\nu$ are directly used at the beginning of the variational procedure.
We see that \eqref{alala} and \eqref{ccmcmc} differ by the term $a \tilde{p}^\rho \tilde{\varpi}_{\rho \nu }$, namely a Generalised Iordanskii force with coefficient $a$. 

This difference arising in the total transverse force on a vortex, is a particular manifestation of a general feature of Carter's multifluid formalism: as \citet{Carter_Starting_Point} pointed out, different choices of chemical basis lead to different non-dissipative hydrodynamic equations, and some important consequences of this fact have been recently discussed also in the context of radiation hydrodynamics \cite{GavassinoRadiazione}.

The formal  implications of this problem can be safely neglected in all the situations in which it is possible to fix a physically motivated gauge choice suggested by the presence of a natural, or particularly convenient, chemical basis. 
In the inner crust case, however, an objective criterion to fix the chemical gauge does not exist: it is therefore  necessary to formulate the hydrodynamic equations in a manifestly gauge-covariant form. This will ensure that the hydrodynamic  model will be valid whatever the microscopic interpretation of $p^\nu$ and $n^\nu$ is, as long as it is compatible with the physical constraints \eqref{barbie}, \eqref{semplifico} and \eqref{joseph}. 
In practice, this means that the equations of the hydrodynamic model must be formulated in such a way as to not change under different chemical gauge fixing choices \cite{Termo}: after the gauge transformation their form should remain exactly the same, but ``with a tilde'' on top of every quantity.

It is apparent that the only way to make \eqref{pulitoeuler} chemical gauge covariant consists of introducing a counter-term $f^{(2)}$ which absorbs the force $a \tilde{p}^\rho \tilde{\varpi}_{\rho \nu}$ generated by the change of chemical basis, see equation \eqref{alala}. Formally, we need to modify \eqref{pulitoeuler} into
\begin{equation}
(n^\rho +\mathcal{Z}p^\rho) \varpi_{\rho \nu} =0 \, ,
\end{equation}
so that, after a change of chemical basis, we are left with
\begin{equation}
(\tilde{n}^\rho +\tilde{\mathcal{Z}}\tilde{p}^\rho) \tilde{\varpi}_{\rho \nu} =0 \, ,
\end{equation}
where
\begin{equation}\label{ZZZZZZZ}
\tilde{\mathcal{Z}}= (1-a)\mathcal{Z}+a \, .
\end{equation}
In the case in which $\mathcal{Z}$ is a constant (and only in this case), the Generalised Iordanskii force can be cancelled out with an appropriate chemical gauge fixing, namely by choosing 
\begin{equation}
\label{ZZZZZZZa}
 a= \dfrac{\mathcal{Z}}{\mathcal{Z}-1} \quad \Rightarrow \quad  \tilde{\mathcal{Z}}=0 \,  .
\end{equation}
%
%
In conclusion, we have proved that the Generalized Iordanskii force is a necessary element to ensure the chemical gauge covariance of the theory.

\subsection{Gauge covariance of the PEVM and the invariance of the HVBK coefficients}

We conclude this section with an analysis of the transformation law under a change of chemical gauge of all the kinetic coefficients, in particular the HVBK coefficients $\slashed{\alpha}$ and $\slashed{\beta}$, we have introduced so far. 
To do this, let us start with the effective neutron density $n^T_{\text{eff}}$ defined by equation \eqref{neFFF}. The transformation laws \eqref{tildiamo} and \eqref{ZZZZZZZ} give
\begin{equation}
 n^T_{\text{eff}} = \tilde{n}^T_{\text{eff}}  \, ,
\end{equation}
which tells us that $n^T_{\text{eff}}$ does not change for different choices of chemical gauge fixing.
Now, let us focus on the generic PEVM: the left-hand side of \eqref{persempresarai} is clearly gauge invariant. We need to ensure the gauge invariance of the right-hand side by imposing
\begin{equation}
\mathcal{R}' n^\nu + \mathcal{R} n^T_{\text{eff}} u^\nu = \tilde{\mathcal{R}}' \tilde{n}^\nu + \tilde{\mathcal{R}} n^T_{\text{eff}} u^\nu \, , 
\end{equation}
so that
\begin{equation}
\mathcal{R}' =\tilde{\mathcal{R}}'   \spc  \mathcal{R} =  \tilde{\mathcal{R}}- \dfrac{a \tilde{p}}{ n^T_{\text{eff}}} \mathcal{R}' \, .
\end{equation}
These transformations also imply  that the effective drag coefficient in \eqref{Reff} is gauge invariant,
\begin{equation}
\mathcal{R}_{\text{eff}}  = \tilde{\mathcal{R}}_{\text{eff}} \, .
\end{equation}
We have shown that the coefficients $n^T_{\text{eff}}$, $ \mathcal{R}'$ and $\mathcal{R}_{\text{eff}}$ do not change under chemical gauge transformations. This immediately tells us that also the HVBK coefficients $\slashed{\alpha}$ and $\slashed{\beta}$ are invariant (i.e. are uniquely defined, in the sense that have the same value regardless of the choice of chemical gauge), namely
\begin{equation}
\slashed{\alpha} =\tilde{\slashed{\alpha}} \spc \slashed{\beta} =\tilde{\slashed{\beta}} \, .
\end{equation}
This is also consistent with the explicit form of the vortex velocity \eqref{lacentrale}, as it is possible to check that all the four vectors $u_v$, $u$, $J$, $l$ are  chemical gauge invariant.

Finally, we also note that it is possible to rewrite the entropy production equation \eqref{secondBoss} and the generic PEVM \eqref{persempresarai} in the manifestly gauge-invariant form
\begin{equation}
    \begin{split}
        & \Theta \nabla_\nu s^\nu = \varpi_{\nu \rho} J^\nu u^\rho \\
        & (n^T_{\text{eff}} u^\rho +J^\rho)\varpi_{\rho \nu} = k \mathfrak{N} {\perp}_{\rho \nu}(\mathcal{R}' J^\rho + \mathcal{R}_{\text{eff}} n^T_{\text{eff}} u^\rho) \, . \\
    \end{split}
\end{equation}
The first relation is simply equation (44) of \citet{GusakovHVBK}. The second relation reveals the deep meaning of the coefficients $n^T_{\text{eff}}$, $ \mathcal{R}'$ and $\mathcal{R}_{\text{eff}}$ and the origin of their chemical-gauge invariance. In fact, we see that they are the prefactors that appear naturally when we rewrite the PEVM in terms of the four manifestly gauge-invariant covectors $u^\rho \varpi_{\rho \nu}$, 
$J^\rho \varpi_{\rho \nu}$, ${\perp}_{\rho \nu} u^\rho$ and ${\perp}_{\rho \nu} J^\rho$.

\section{The interpretation of the Generalised Iordanskii force}
\label{sezione5}

At the mesoscopic scale, the flow past a vortex in a simple perfect fluid (or in a superfluid at zero temperature) gives rise to a transverse Magnus force that is given by the well known Joukowski lift formula. 
The problem of generalising this to multiconstituent superfluid models has been controversial since it was originally posed by the work of Iordanskii in the context of the Landau two-fluid model for $^4$He \cite{donnelly_book,sonin_book_2016,sourie2020IJMPB}. 

From a geometrical perspective, this problem has been fully clarified by \citet{Carter_Prix_Magnus} for a  generic relativistic multifluid. An analogous analysis has been recently proposed for the Newtonian $^4$He case at the mesoscopic scale~\cite{sourie2020IJMPB}.
The analysis of \cite{Carter_Prix_Magnus}  can be used to provide an interpretation to $\mathcal{Z}$ in terms of the circulation of the momentum $\chi_\nu$ around a single vortex. 
This allows us to reinterpret the long-lasting controversy on the presence of a Iordanskii force acting on quantized vortices in view of  the chemical gauge invariance of the mesoscopic hydrodynamic model.


\subsection{Forces on a vortex at the mesoscopic scale: Carter's multifluid approach}

Let us consider again the generic PEVM \eqref{persempresarai}: as discussed in Sec \ref{forrRza!}, if we divide it by the vortex density $\mathfrak{N}$, we can interpret the resulting equation as the request that the  forces per-unit-length acting on a single vortex line should balance. 
It is natural to interpret the force appearing on the left-hand side of \eqref{persempresarai} as the multifluid analogue of the total Joukowski lift force per unit length $\mathcal{F}^J_\nu$, namely 
\begin{equation}
\label{wdieokvmok}
\mathfrak{N} \mathcal{F}^J_\nu = (n^\rho +\mathcal{Z}p^\rho) \varpi_{\rho \nu} \, ,
\end{equation} 
 and the one on the right-hand side (given its dissipative character) as the total drag force acting on a vortex (in fact, according to \eqref{ssssSsecondddLav}, it contributes to the entropy increase). 
Note that, according to our definition, the total Joukowski lift force represents the totality of the transverse forces acting on a vortex, see \eqref{persempresarai}. 
 
\citet{Carter_Prix_Magnus} computed the  Joukowski lift in full generality by studying the momentum balance on the flows around a vortex segment, but then they assumed the validity of the mesoscopic model  introduced in Sec \ref{sezione2}. In fact, they imposed the zero vorticity condition \eqref{zerovort} as the equation which closes the system, and studied the momentum flux through a closed loop surrounding the vortex (the loop was taken sufficiently far from the vortex core to guarantee the reliability of the hydrodynamic description at the mesoscopic scale, see also the discussion in~\cite{sourie2020IJMPB}), obtaining  
\begin{equation}
\label{FFJJ}
\mathcal{F}^J_\nu =\dfrac{1}{k \mathfrak{N}} \varpi_{\rho \nu} \sum_X \mathcal{C}^X n_X^\rho,
\end{equation}
see equation (21) of \cite{Carter_Prix_Magnus}. 
Here, the label $X$ labels the currents of the model and
\begin{equation}
\label{CCCCC}
\mathcal{C}^X = \oint \mu^X_\sigma dx^\sigma
\end{equation}
is the circulation integral of the momentum conjugate to current $X$ along the loop (which they prove not to depend on the chosen loop, as long as it goes around a single vortex once). 
We remark that in equation \eqref{FFJJ} the currents $n_X^\nu$ should be interpreted as the large-scale ones, namely those used in the macroscopic model, while in equation \eqref{CCCCC} the momentum is the mesoscopic one, measured at the inter-vortex separation scale \cite{Carter_Prix_Magnus}.
In our case, where we have only the three currents $s^\rho$, $p^\rho$ and $n^\rho$, equation \eqref{FFJJ} explicitly reads 
\begin{equation}\label{paperibo}
\mathfrak{N} \mathcal{F}^J_\nu = \dfrac{1}{k} \varpi_{\rho \nu} (\mathcal{C}^s s^\rho + \mathcal{C}^p p^\rho + \mathcal{C}^n n^\rho),
\end{equation}
with
\begin{equation}\label{circolazioniIIIIIII}
\mathcal{C}^s =\oint \Theta u_\sigma dx^\sigma  \quad \quad \mathcal{C}^p = \oint \chi_\sigma dx^\sigma  \quad \quad   \mathcal{C}^n = \oint \mu_\sigma dx^\sigma  = k = \pi \hbar \, ,
\end{equation}
see equation \eqref{joseph}.
Since $\mathcal{C}^s s^\rho$ is of the second order in $\Theta$, we can neglect it in the low temperature limit. 
Moreover, assuming that each vortex contains a single quantum of circulation, the Josephson relation \eqref{joseph} implies the Feynman-Onsager quantization relation $\mathcal{C}^n=k$, which immediately tells us that $\mathcal{C}^n$ is chemical gauge independent (i.e. it is a physically relevant quantity).  
Now, equation \eqref{paperibo} becomes
\begin{equation}
\mathfrak{N} \mathcal{F}^J_\nu \approx  \varpi_{\rho \nu} \bigg( n^\rho + \dfrac{\mathcal{C}^p}{k} p^\rho  \bigg) 
\end{equation}
and direct comparison with \eqref{wdieokvmok} provides us with the mesoscopic interpretation of $\mathcal{Z}$, 
\begin{equation}\label{fnvjifrv}
\mathcal{Z} = \dfrac{\mathcal{C}^p}{k} \, .
\end{equation}
Finally, as a consistency check, we can re-derive the affine transformation law \eqref{ZZZZZZZ} of $\mathcal{Z}$   directly from \eqref{fnvjifrv}:
employing the first equation in \eqref{tuttiinbarca}, we find
\begin{equation}\label{cnonc}
\tilde{\mathcal{C}}^p = \oint \tilde{\chi}_\sigma dx^\sigma = (1-a)\mathcal{C}^p +ak \, ,
\end{equation}
and by dividing both sides by $k$ we recover equation \eqref{ZZZZZZZ}.

The Iordanskii controversy, transported to the neutron-star setting, revolves around the problem of prescribing a value for $\mathcal{C}^p$. Different assumptions about the mesoscopic behaviour of the fluid lead to different formulas for $\mathcal{C}^p$ and therefore to a different predicted intensity of the force $f^{(2)}$ (the transverse force proportional to $p^\rho \varpi_{\rho \nu}$). In the remaining part of the section, we summarise the most common available prescriptions for $\mathcal{C}^p$, recalling the mesoscopic assumptions that lead to these formulas.

\subsection{Transverse forces on a vortex: the Sonin-Stone model}

As anticipated in section \ref{sezione4}, the original hydrodynamic model of \citet{langlois98} is not based on the general PEVM of the form \eqref{generaleBuf}, or \eqref{persempresarai}, but rather the force balance in the surroundings of a vortex is governed by the ``standard'' PEVM in \eqref{nuvarpi}. To obtain this standard prescription for vortex dynamics, one needs to impose that
\begin{equation}
\label{uccidic}
\mathcal{C}^p=0 \, ,
\end{equation}
which immediately implies that one should work in the particular chemical gauge such that 
\begin{equation}
n^T_{\text{eff}} =n^T \, .
\end{equation}
\citet{Carter_Prix_Magnus} noticed  that the prescription \eqref{uccidic} of setting the circulation of the normal momentum to zero is the same assumption that leads in models for $^4$He to the PEVM of \citet{sonin1997PhRvB} and \citet{stone2000}, see also \cite{Sonin_comment_1998,sonin_book_2016}.
The same conclusion, namely that assuming no circulation of the normal fluid \emph{momentum} around a vortex leads to the forces on a vortex derived by Sonin and Stone, was also reached by \citet{Geurst1997} in the context of the non-relativistic two-fluid model for $^4$He. 

If, for some mesoscopic reason, the condition \eqref{uccidic} holds for a vortex immersed in a neutron star crust, then the total Joukowski lift force $\mathfrak{N} \mathcal{F}^J_\nu$ and the Carter-Magnus force 
$ n^\rho \varpi_{ \rho \nu}$ should coincide, see \eqref{wdieokvmok}. 
However, given the transformation law \eqref{cnonc}, equation \eqref{uccidic} could be valid only in a particular chemical gauge and, in general, it will not hold for any other choice of chemical gauge fixing. Therefore, we are left with the problem of identifying the preferred chemical basis for which $\mathcal{C}^p=0$.

In applying their model (that is in principle valid also in the outer core) to the inner crust dynamics, \citet{langlois98} did not specify in which chemical basis they were working. They only introduced the formal assumption that it was possible to split the total baryon current into a current of superfluid neutrons and a collectively comoving normal part. As a working hypothesis, one may assume that the preferred chemical basis is the one in which $n^\nu$ counts all the neutrons (the dripped ones, as well as the ones confined in nuclei). Thus, by imposing \eqref{nuvarpi}, one would implicitly assume that \eqref{uccidic} holds for this particular choice. 

This, however, looks quite unnatural. 
From a practical perspective it makes sense to expect that the neutrons that are confined on cosmological timescales only play the role of additional mass carried by the effective normal component, so that one would expect that these confined neutrons should be included in the currents $p^\nu$, and, hence, contribute to the normal momentum $\chi_\nu$. 
Clearly, even accepting this qualitative argument, we would still be left with the question of which mass number $A$ to impose, see subsection \ref{chemicalGaugeEEEEE}. Therefore, the problem of whether setting $\mathcal{C}^p=0$ is justified remains open.

Interestingly, there is an alternative way to extend the results of \citet{sonin1997PhRvB} and \citet{stone2000} to the neutron star context. In fact, \citet{Geurst1997} and \citet{sourie2020IJMPB} have shown that if we match the phonon scattering models with asymptotic two-fluid hydrodynamics (far from the vortex core) then we obtain
\begin{equation}\label{kinetoBob}
 \oint u_\sigma dx^\sigma \approx \dfrac{k}{\mu_T}.
\end{equation}
This statement is clearly chemical-gauge invariant. In App \ref{vbaf} we show that this assumption, applied to the context of neutron-star crusts, would lead to the chemical-gauge covariant prescription\footnote{The reason why equations \eqref{uccidic} and \eqref{hjklm} are so different, while in the $^4$He analogue they coincide, is that in helium the formal analogue of the relativistic chemical potential of the normal component is the temperature $\Theta$, while the formal analogue of the relativistic chemical potential of the superfluid component is $\approx mc^2$. Therefore in equation \eqref{hjklm} one deals with the ratio $\Theta/(mc^2)$, which in the Newtonian limit is zero.}
\begin{equation}\label{hjklm}
    \mathcal{C}^p =  k \, \dfrac{\chi_T}{\mu_T}  \spc  \chi_T = - \chi_\nu u^\nu.
\end{equation}
Assuming that the deviations from beta equilibrium are small ($\chi_T \approx \mu_T$), we immediately obtain
\begin{equation}
n^T_{\text{eff}} = b^T  \spc b^T =-b^\nu u_\nu \, ,
\end{equation}
and therefore
\begin{equation}\label{SoninProprioLui}
\mathfrak{N} \mathcal{F}^J_\nu = b^\rho \varpi_{\rho \nu} \, .
\end{equation}
This formula for the Joukowski lift force is clearly chemical-gauge invariant and can be considered to be the natural generalization of the result  of Sonin and Stone to the neutron star crust case. 

We remark, however, that the calculations of Sonin and Stone are performed in the specific context of bosonic superfluidity (as they are based on phonon dynamics) and that the results of \cite{Carter_Prix_Magnus} are valid only in the absence of viscosity and elasticity, which in neutron-star crusts, at the mesoscopic scale, can become very important. Thus, equation \eqref{SoninProprioLui} constitutes more a philosophical analogy, than a reliable formula. Improving it is beyond the scope of the present paper. 

Note also that the present discussion cannot be extended to the outer core, where the medium is expected to be homogeneous (i.e. there is no ambiguity in the operational definition of the ``free'' neutrons) and the protons are superconducting. These two physical conditions break the chemical-gauge covariance of the hydrodynamic description, and the circulation of the momentum associated with the protons around a neutron vortex assumes the well defined value $\mathcal{C}^p=0$, unless the vortex line is surrounded by (or superimposed to) one or more flux-tubes and the loop embraces both kinds of topological defects \cite{alpar_up_84,mendell1991,andersson_sidery_2006,sourie_corepinn_2020MNRAS}. 
Thus, in the core one should follow a prescription for the lift force like the one presented in subsection \ref{eulero}.

\subsection{Transverse forces on a vortex: the Thouless-Ao-Wexler-Geller model}

There is an other simple assumption for the value of $\mathcal{C}^p$, alternative to \eqref{uccidic} and \eqref{hjklm}, which makes the Joukowski lift force \eqref{wdieokvmok} manifestly chemical gauge invariant. 
To explore  this alternative possibility, let us first introduce the gauge-invariant quantity $Y^{-1}$, that is closely  related to the relativistic generalization of the usual superfluid density appearing in the Landau two-fluid model~\cite{Gusakov2007,Termo}.

The conjugate momenta $\chi_\nu$ and $\mu_\nu$ can always be uniquely expressed as linear combinations of the two currents $p^\nu$ and $n^\nu$,
\begin{equation}
\label{decompbubu}
\chi_\nu = \mathfrak{B}\,  p_\nu +\mathfrak{A} \, n_\nu  \spc  \mu_\nu = Y^{-1} n_\nu + \mathfrak{A}\, p_\nu \, ,
\end{equation}
where the same coefficient $\mathfrak{A}$ appears in both expressions to guarantee the symmetry of the stress-energy tensor \eqref{tinuro}, see e.g. \cite{Carter_Starting_Point,Termo}.
Under a chemical gauge transformation, the  coefficients change as\footnote{
Insert \eqref{decompbubu} into \eqref{tuttiinbarca} and  employ \eqref{tildiamo}.
The third equation in \eqref{qpsmodlx} tells us that $Y$ is indeed chemical gauge invariant. 
}
\begin{equation}\label{qpsmodlx}
\begin{split}
& \tilde{\mathfrak{B}}=(1-a)^2 \mathfrak{B}+2a(1-a)\mathfrak{A}+a^2 Y^{-1} \\
& \tilde{\mathfrak{A}}=(1-a)\mathfrak{A}+a Y^{-1} \\
& \tilde{Y}=Y \, . \\
\end{split}
\end{equation}
To complete the link with the formalism used in  \cite{Gusakov2007,Gusakov2016,Termo}, let us decompose the superfluid momentum in the frame of the normal component, similarly to what has been done in \eqref{pizzobaldo}, namely
\begin{equation}
\mu_\nu = \mu_T u_\nu +w_\nu  \spc w_\nu u^\nu =0 \, .
\end{equation}
It is easy to verify that \eqref{decompbubu} implies the chemical gauge invariant relation
\begin{equation}
J^\nu = Y w^\nu \, ,
\end{equation}
which is the definition of the entrainment coefficient of \citet{Gusakov2007}, see equations (3) and (27) therein.

Now, the relations in  \eqref{qpsmodlx} allow to derive the transformation law
\begin{equation}
k\tilde{Y}\tilde{\mathfrak{A}} = (1-a)kY\mathfrak{A} +ak \, .
\end{equation}
A comparison with \eqref{cnonc} gives that $k\tilde{Y}\tilde{\mathfrak{A}}$ transforms exactly  as $\mathcal{C}^p$. Therefore, if, instead of \eqref{uccidic} or \eqref{hjklm}, it is rather assumed that
\begin{equation}
\label{thouless}
\mathcal{C}^p = k Y \mathfrak{A} \, ,
\end{equation}
we would end up with a chemical gauge covariant prescription for the value of $\mathcal{C}^p$. With this assumption the Joukowski lift force becomes
\begin{equation}
\label{thouless2}
\mathfrak{N} \mathcal{F}^J_\nu = Y \mu^\rho \varpi_{\rho \nu} \, ,
\end{equation}
which is gauge-invariant as well (the right hand side is a manifestly gauge-invariant quantity). 
Finally, the effective neutron density introduced in \eqref{neFFF} is
\begin{equation}
n^T_{\text{eff}} = Y \mu_T \, ,
\end{equation}
which is the (chemical gauge invariant) Landau density of superfluid neutrons \cite{carter92}.

In App \ref{riGGG} we show that equation \eqref{thouless} can be obtained from mesoscopic considerations if one assumes that, as a result of the action of fast dissipative processes, the fluid has reached thermodynamic equilibrium in a neighbourhood of the vortex (which is a chemical gauge invariant statement). 
The calculation is a rigorous proof of the argument of \citet{Carter_Prix_Magnus}, who have proposed that assuming that the normal component is rigid might lead to the result of Thouless and collaborators for the transverse forces on a vortex. 
Indeed, this was one of the central assumptions invoked by \citet{TAN_original1996} in the preliminary microscopic analysis which led \citet{Wexler1997} to formulate the Newtonian version of \eqref{thouless2}. However, to make the argument rigorous, one needs also to assume diffusive equilibrium.

Contrarily to \eqref{SoninProprioLui}, equation \eqref{thouless2} does not depend on the microscopic details of the system and is not affected by the inclusion of elasticity and viscosity. This is due to the fact that it is based only on the general properties of the thermodynamic equilibrium state.

Interestingly, equation \eqref{thouless2} has been implicitly postulated by \citet{GusakovHVBK} to obtain his formulation of the no-drag limit of the HVBK hydrodynamics,
\begin{equation}
\mu^\rho \varpi_{\rho \nu}=0 \, ,
\end{equation}
see equation (60) of \cite{GusakovHVBK}.

\subsection{The Iordanskii force in neutron stars}

We are finally able to transport the standard Iordanskii problem, well known in superfluid $^4$He, to the context of neutron star crusts. 

To do this we, first of all, invert the second equation of \eqref{decompbubu} as follows,
\begin{equation}\label{defnvkfm}
n^\rho = Y \mu^\rho - Y\mathfrak{A} p^\rho.
\end{equation}
The two terms in the right-hand side are the analogue, for a relativistic mixture, of the Landau superfluid part $n_S^\nu$ (which is chemical gauge invariant) and the Landau normal part (which is chemical gauge dependent) $n_N^\nu$ of the current $n^\nu$,
\begin{equation}
n_S^\rho = Y \mu^\rho  \spc n_N^\rho = - Y\mathfrak{A} p^\rho,
\end{equation}
see also \cite{carter92}. These can be used to decompose the total baryon current into a superfluid and a normal part (in the Landau sense) 
\begin{equation}\label{baRYUNZ}
    b^\rho = b_S^\rho + b_N^\rho  \spc b_S^\rho = n_S^\rho = Y \mu^\rho  \spc b_N^\rho =p^\rho+ n_N^\rho =(1 - Y\mathfrak{A}) p^\rho   .
\end{equation}
This decomposition is chemical-gauge invariant and therefore has an unambiguous physical meaning.

Contracting \eqref{baRYUNZ} with $\varpi_{\rho \nu}$ we obtain
\begin{equation}
f^{S}_\nu = f^T_\nu + f^I_\nu,
\end{equation}
where
\begin{equation}
f^{S}_\nu = b^\rho \varpi_{\rho \nu}
\end{equation}
is the neutron star analogue of the Sonin force \eqref{SoninProprioLui}, while
\begin{equation}
f^T_\nu = b_S^\rho \varpi_{\rho \nu}=Y \mu^\rho \varpi_{\rho \nu}
\end{equation}
is the neutron star analogue of the Thouless force \eqref{thouless2} and
\begin{equation}\label{LaIordainfdiemnticalbule}
f^I_\nu = b_N^\rho \varpi_{\rho \nu}= (1- Y\mathfrak{A}) p^\rho \varpi_{\rho \nu}
\end{equation}
is the neutron star analogue of the Iordanskii force (which, as we see, is a particular case of Generalised Iordanskii force).

The controversy, in its standard formulation, revolves around the presence of $f^I$ in the total Joukowski lift force, with Sonin and collaborators who argue that it should be included, and Thouless and collaborators who argue that it should be removed.

\section{ Towards a resolution of the Iordanskii force controversy }
\label{sec6}

In this final section we present a possible resolution of the controversy surrounding the Iordanskii force \cite{Sonin_comment_1998}: the two incompatible assumptions $\mathcal{C}^p = k \chi_T/\mu_T$ (which gives the transverse force on a vortex of Sonin and Stone) and $ \mathcal{C}^p = k Y \mathfrak{A} $ (leading to the transverse force of Thouless and collaborators) possibly refer to two different dynamical regimes of the same system. 

Our argument is based on a revised version of the thought experiment of \citet{Wexler1997}: 
with the aid of the thermodynamic tools devised in \cite{Termo} and \cite{GavassinoTermometri},
we will  extend the original experiment to an arbitrary relativistic superfluid-normal mixture. Then, we will show that, depending on the time-scale at which Wexler's experiment is performed, one can move from the case in which  $ \mathcal{C}^p = k Y \mathfrak{A} $, that we will call the \textit{Thouless regime}, to a situation in which $ \mathcal{C}^p = k \chi_T/\mu_T $, the \textit{Sonin-Stone regime}.

Let us remark that the arguments presented in this last section transcend the specific interest for neutron stars. 
In fact, given the present, simplified and effective, two-fluid description of neutron star crusts (that is formally analogous to the one of \cite{GusakovHVBK} for a  single-species superfluid at finite temperature), our reasoning applies also to $^4$He.

\subsection{ Geometry and preliminary definitions of Wexler's gedanken experiment }
\label{lointrodescribo}

The original experiment of \citet{Wexler1997} aims to compute the Joukowski lift force exerted on a vortex during a quasi-static transformation by means of a thermodynamic argument (the quasi-static assumption is central, so that dissipative effects due to vortex motion are negligible).

\begin{figure}
    \centering
    \includegraphics{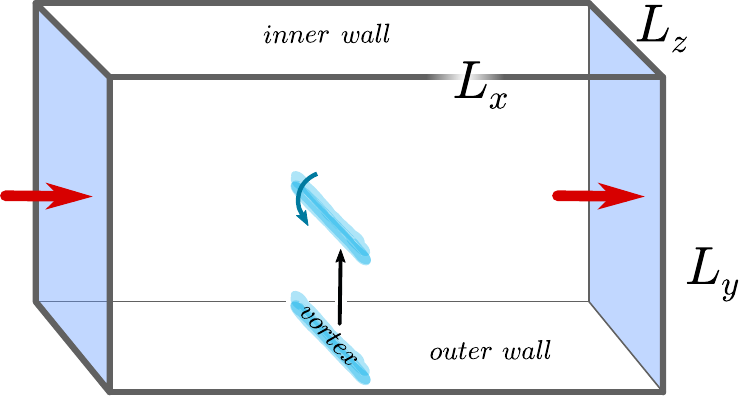}
    \caption{Sketch of the annulus  considered by \citet{Wexler1997}: periodic boundary conditions are imposed on the two shaded faces, that have no walls. The two red arrows indicate the direction of the uniform background flow along the circumference $L_x$ (since $L_x \gg L_y$, this background flow is almost uniform, at least far from the vortex). An intermediate displacement of the vortex (that from the outer wall is slowly dragged towards the inner wall) is also shown. The vorticity is directed out of the page, in accordance with the sketched direction of the induced flow around the vortex. }
    \label{fig_wexler}
\end{figure}

Let us consider a portion of our two-fluid system enclosed in a (non-rotating) annulus with rectangular section (for simplicity, the spacetime is assumed flat). Adopting the same notation of Wexler, we call $L_x$ the circumference of the annulus, $L_y$ the basis of the rectangular section and $L_z$ its height, with $L_y \ll L_x$, so that all the hydrodynamic quantities can be considered uniform along the section, see Fig \ref{fig_wexler}. 
The walls play the role of a heat bath for the fluid, meaning that they have both infinite heat capacity and inertial mass \citep{GavassinoTermometri}.

The experiment proceeds as:
\begin{enumerate}
\item In the initial state, the total system (the fluid and the walls) is in global thermodynamic equilibrium, with a winding number of the superfluid phase $\phi$ along $L_x$ equal to $N$ (i.e. $\mu_x L_x = k N)$. In fact, the winding number is one of the thermodynamic variables that should be specified to identify a well defined equilibrium state \cite{Termo}.
\item A vortex is created on the outer boundary of the annulus and slowly dragged by an external force   to the inner boundary where it is, then, annihilated. 
\item In the final state the total system is, again, in global thermodynamic equilibrium, but with a winding number of the  phase $\phi$  equal to $N+1$.
\end{enumerate}

By studying the energetics of this experiment (we use the relativistic thermodynamic formalism formulated in \cite{GavassinoTermometri,Termo}), it is possible to extract the total lift force exerted by the fluid on the vortex during the process.

\subsection{Relativistic thermodynamic analysis}

The description of the aforementioned experiment in a relativistic setting requires the use of a relativistic thermodynamic formalism. We will adopt the formulation proposed by \citet{GavassinoTermometri}, where the concept of work is extended in special relativity as  
the  variation of the total four-momentum (in the Newtonian limit the usual concept of work is recovered).

The fluid in the annulus has total four-momentum $\mathcal{P}^\nu$ and total entropy $S$, while the 
heat bath has total four-momentum $\mathcal{P}_H^\nu$ and total entropy $S_H$. 
Therefore, the  four-momentum and the entropy of the total system are 
\begin{equation}
\mathcal{P}_{\text{tot}}^\nu = \mathcal{P}^\nu+ \mathcal{P}^\nu_H  \spc 
S_{\text{tot}} = S+S_H \, . 
\end{equation}
The external force that drags the vortex can be modelled as the result of the presence of an external field in the microscopic Lagrangian for the total system, which breaks the invariance of the theory under the full Poincar\'{e} group and, hence, also the conservation of $\mathcal{P}_{\text{tot}}^\nu $, that changes by a quantity $\delta \mathcal{P}_{\text{tot}}^\nu $ during the whole process. 
Following \cite{GavassinoTermometri}, the work four-vector  made by the external force between the beginning and the end of the process is, therefore,
\begin{equation}\label{worKKK}
\delta \mathcal{W}^\nu = \delta \mathcal{P}_{\text{tot}}^\nu  
= \delta \mathcal{P}^\nu+ \delta \mathcal{P}^\nu_H \, .
\end{equation}
In addition, since the process is assumed to be infinitely slow, we can impose that no dissipation occurs, namely (the system fluid+annulus+external field is isolated and the entropy of the external field is zero)
\begin{equation}\label{nondiss}
\delta S+ \delta S_H =0 \, .
\end{equation}
Defining the mass and the center-of-mass four-velocity of the heat bath as  
\begin{equation}
M_H = \sqrt{-\mathcal{P}_H^\sigma \mathcal{P}_{H\sigma}}  \spc
u_H^\sigma = \dfrac{\mathcal{P}_H^\sigma}{M_H} \, ,
\end{equation}
we impose, from covariance requirements, that its equation of state is given in the form
$ S_H =S_H(M_H) \,$, whose differential is
\begin{equation}\label{dsssh}
\delta S_H = - \dfrac{u_{H\, \sigma}}{\Theta_H} \, \delta \mathcal{P}_H^\sigma \, ,
\end{equation}
where we have defined the temperature of the heat bath through the relation
\begin{equation}
\dfrac{1}{\Theta_H} = \dfrac{dS_H}{dM_H} \, .
\end{equation}
Contracting both sides of \eqref{worKKK} with $u_H^\nu$ and employing equations \eqref{dsssh} and \eqref{nondiss}, we obtain
\begin{equation}
-u_{H \, \sigma} \, \delta \mathcal{W}^\sigma = 
-u_{H\, \sigma} \, \delta \mathcal{P}^\sigma - \Theta_H \, \delta S \, .
\end{equation}
At this point, the free energy of the fluid is defined as
\begin{equation}
F = -u_{H \, \sigma} \,  \mathcal{P}^\sigma - \Theta_H  S \, ,
\end{equation}
and we find that (in the limit of infinite mass and heat capacity of the heat bath)
\begin{equation}\label{freeBalance}
-u_{H\nu} \delta \mathcal{W}^\nu = \delta F \, .
\end{equation}
This is the relativistic analogue of the well-known Newtonian theorem which states that the work exerted on a system in a slow isothermal process coincides with the change of its free energy \cite{GavassinoTermometri}. Now, following the same steps of \citet{Wexler1997}, we need to calculate explicitly both sides of equation \eqref{freeBalance}. 

\subsection{Computation of the variation of the total four-momentum}

From now on we will work in the inertial frame of the walls, where the time coordinate is $t$, 
and  $0$ and $t_{\text{end}}$ are the initial and the final instants of the slow process.
The work four-vector in  \eqref{worKKK} can be written more explicitly as
\begin{equation}
\delta \mathcal{W}^\nu = \int \bigg( \nabla_\rho T^{\rho \nu}+ \nabla_\rho T^{\rho \nu}_H \bigg) d^4 x
\, ,
\end{equation}
where the integral extends over the whole spacetime region in which the process occurs and $T^{\rho \nu}_H$ is the stress-energy tensor of the heat bath. 
Clearly, wherever the external force is not applied
\begin{equation}
\nabla_\rho T^{\rho \nu}+ \nabla_\rho T^{\rho \nu}_H  =0 \, ,
\end{equation}
which means that we can restrict our integration volume to a moving spatial domain $\mathcal{K}(t)$ surrounding the vortex line. 
Assuming that $\mathcal{K}(t)$ do not extend into the walls for $0<t<t_{\text{end}}$, we have that $T^{\rho \nu}_H=0$ inside $\mathcal{K}(t)$, implying
\begin{equation}
\delta \mathcal{W}^\nu = \int_0^{t_{\text{end}}} \int_{\mathcal{K}(t)} \nabla_\rho T^{\rho \nu}
\,  d^3 x \, dt \, . 
\end{equation}
In the quasistatic approximation we have
\begin{equation}\label{wodkmlsx}
\int_{\mathcal{K}(t)} \nabla_\rho T^{\rho j} d^3 x = \int_{\partial \mathcal{K}(t)}  T^{\rho j} 
\, d^2 \Sigma_\rho \, ,
\end{equation} 
where $\partial \mathcal{K}(t)$ is the boundary of $\mathcal{K}(t)$ and $d^2 \Sigma_\rho$ is its normal covector.  
The right-hand side of \eqref{wodkmlsx} is nothing but the total  Joukowski lift force $\mathcal{F}^{Jj}$ for a stationary vortex configuration (integrated over the vortex line), see e.g. \cite{Carter_Prix_Magnus}. 
Therefore, since the fluid is essentially homogeneous in the annulus (far enough from the vortex filament), the lift force is constant during the whole process (apart in the initial and final moments, where the vortex feels the effect of the walls), leading us to the final formula
\begin{equation}
\delta \mathcal{W}^j =L_z \, \mathcal{F}^{Jj} \, t_{\text{end}} \, .
\end{equation}
Unfortunately, the same kind of explicit calculation cannot be made for $\delta \mathcal{W}^0$, due to a problem of infinities. In fact, as the process is carried out slower and slower to reach the reversible limit, the integration in time diverges, while the integral in space converges to $\mathcal{F}^{J0} =0$, giving an indeterminate result of the kind $ \infty \times 0$. 
To solve this drawback, let us focus on the source of the external force that drags the vortex. 

Following \citet{landau5}, we can model the external force as the result of the interaction of the system with an external machine (the external field we mentioned earlier) that pins the vortex and drags it. Since this machine should drag the vortex without altering its own state of motion relevantly, we can assume that it has an effectively infinite inertia $M_{\text{e}}$, and its four-momentum is given by 
\begin{equation}
\mathcal{P}^\sigma_{\text{e}} \, = \, M_{\text{e}} \,  u_{v}^\sigma \, , 
\end{equation}
where $u_{v}^\sigma$ is the four-velocity with which the pinned vortex is dragged. 
Since the system fluid+machine+heat bath is isolated, we can impose
\begin{equation}
\label{labbo}
\delta \mathcal{P}^\sigma_{\text{e}}=-\delta \mathcal{W}^\sigma \, .
\end{equation}
Now, we need to treat the machine as a purely mechanical macroscopic body with no internal degrees of freedom and set its entropy to zero \cite{landau5}. Hence, its equation of state reduces to the material-particle relation $M_{\text{e}} = \text{const}$, which   implies
\begin{equation}\label{labbu}
\delta \mathcal{P}^\sigma_{\text{e}}= M_{\text{e}} \, \delta u_{v}^\sigma \, .
\end{equation}
Combining \eqref{labbo} and \eqref{labbu} we immediately derive the geometric constraint
\begin{equation}\label{dcfv}
\delta \mathcal{W}_\sigma \,  u_{v }^\sigma=0 \, .
\end{equation} 
In the reference frame of the walls we can write 
$u_v=\Gamma_v(1, \Delta^1, \Delta^2, \Delta^3)$, 
which allows us to reformulate equation \eqref{dcfv} as
\begin{equation}\label{odc}
\delta \mathcal{W}^0 = \delta \mathcal{W}_j \, \Delta^j 
= L_z \, t_{\text{end}} \, \mathcal{F}^{J}_j \, \Delta^j \,  .
\end{equation}
We have recovered, in the context of relativistic thermodynamics, the well-known Newtonian expression for the work of a force. As the vortex is dragged along the positive $y$-axis (see Fig \ref{fig_wexler}), we impose
\begin{equation}
\Delta^j = \dfrac{L_y}{t_{\text{end}}} \, \delta\indices{^j _y} \, ,
\end{equation} 
which, plugged into \eqref{odc}, gives
\begin{equation}\label{deltaW0}
\delta W^0 = L_y L_z \, \mathcal{F}^{J}_y \, . 
\end{equation}
This is the formula for the left-hand side of \eqref{freeBalance} we were looking for, and constitutes the relativistic version of equation (5) of \citet{Wexler1997}.

\subsection{Computation of the free-energy variation}

Since both the initial and the final state are in full thermodynamic equilibrium, we can use the prescription of \cite{Termo}, see Sec 3.1 therein, for the variation of the Helmholtz free-energy per unit volume $\mathfrak{F}$ of a multifluid, which  in this case  reads
\begin{equation}
\label{strampalatavariazione}
\delta \mathfrak{F} = -s \delta \Theta + \chi_T \delta p + \mu_T \delta n^T + n^j \delta \mu_j \, .
\end{equation}
This formula is based on the fact that in thermodynamic equilibrium the normal component needs to be at rest in the frame of the heat bath, namely 
\begin{equation}\label{comosso}
u^\nu = u_H^\nu \, .
\end{equation}
To explicitly compute the variation in \eqref{strampalatavariazione}, we start by noticing that in equilibrium $\Theta = \Theta_H$ (where $\Theta_H$ is constant), meaning that $\delta \Theta =0$.
Furthermore, given that no reaction occurs and the baryon number is conserved, the total numbers $N_p = L_x L_y L_z p^0$ and $N_n = L_x L_y L_z n^0$ are constant in every chemical gauge, implying that $\delta p = \delta n^T =0$. From \eqref{comosso} it immediately follows that $p^j=0$, which inserted in the $j$-th component of equation \eqref{defnvkfm}, gives
\begin{equation}\label{nYjm}
n^j = Y \mu^j \, .
\end{equation}
With these additional relations it is finally possible to write down the explicit result for \eqref{strampalatavariazione}: the variation of the total free energy $F=L_x L_y L_z \mathfrak{F}$ reads
\begin{equation}
\delta F = L_x L_y L_z Y \mu^j \delta \mu_j \, . 
\end{equation}
Now, the only missing ingredient is the variation of the superfluid momentum during the process. 
The only non-vanishing component of $ \mu_j $ is the one along the annulus, in the $x$-direction. 
Hence, given the Josephson relation \eqref{joseph} and the fact that the phase has a winding number $N$ in the initial state and $N+1$ in the final state, we obtain
\begin{equation}
\mu_x = N \dfrac{k}{L_x}  \spc \mu_x + \delta \mu_x = (N+1) \dfrac{k}{L_x} \, ,
\end{equation}
which gives
\begin{equation}\label{deltaFFF}
\delta F =  ( L_y L_z) \, Y \, \mu^x \, k \, . 
\end{equation}
This is exactly the relativistic version of equation (11) of \citet{Wexler1997}.

\subsection{Thermalised, intermediate and ballistic regimes}

Combining equations \eqref{freeBalance}, \eqref{deltaW0} and \eqref{deltaFFF}, we find that the outcome of Wexler's experiment predicts a Joukowski lift force given by (remember that the vortex is fixed in the frame of the heat bath, so that the vortex velocity does not appear explicitly)
\begin{equation}
\label{eonfm}
\mathcal{F}^{J}_y = Y \mu^x k \, .
\end{equation}
This formula is exactly the total transverse force acting on a vortex found by \citet{Thouless2001},
see also \cite{geller2000}, and is in agreement with the final result of \cite{Wexler1997}, see equation (14) therein. 
However, it is fundamental to note that to obtain \eqref{eonfm}  we had to use equation \eqref{nYjm}, 
which also implies 
\begin{equation}\label{ikgbmlfv}
\mathcal{F}^{J}_y =  n^x k =b^x k = (n^x+\mathcal{Z}p^x) k \spc  \forall \mathcal{Z} \, ,
\end{equation}
that is in agreement with the result of Sonin and Stone \cite{sonin_book_2016}: the thermodynamic argument of Wexler, as it is formulated, is not sufficient to pin down a unique, well defined, value of $\mathcal{Z}$, in neutron star crusts as well as in $^4$He. 

There is a physical reason for this. Since the transformation described in Sec \ref{lointrodescribo} is quasi-static, the fluid passes trough a sequence of  states that are in thermal equilibrium with the heat bath, meaning that equation \eqref{comosso} is verified for all $t$. 
At the same time, since the vortex is dragged infinitely slowly, we also have
$ u_v^\nu = u_H^\nu =u^\nu$, and from  \eqref{uabaobab} we obtain
\begin{equation}
u^\nu = {\paral \,}\indices{^\nu _\rho} u^\rho \, ,
\end{equation}
which implies
\begin{equation}
p^\rho \varpi_{\rho \nu} =0 \, ,
\end{equation}
during the whole process. This proves that  the exact value of $\mathcal{Z}$ is irrelevant to the energetics of the quasi-static transformation. 
This is also consistent with the original study of \citet{Wexler1997}, who used his experiment to constrain just one of the two free parameters appearing in the total lift force (the so called $A$-parameter, see equation (4) in \cite{Wexler1997}) and then invoked the results of \citet{TAN_original1996} to constrain the other. 
With the aid of more recent results, however, we can now gain further insight from Wexler's thought experiment, which indicates a possible resolution of the controversy on the total transverse lift force acting on a vortex, and that we briefly discuss in the following.

Let us study the behaviour of the normal component during the quasi-static process. 
As we said, since the transformation is infinitely slow, equation \eqref{comosso} holds everywhere. 
Therefore, even if both \eqref{eonfm} and \eqref{ikgbmlfv} are correct, from the practical point of view we are in the scenario described in App \ref{riGGG}, where the normal component can be thought as rigid\footnote{
This argument is valid for any mixture of superfluid and normal species, or single-species superfluids at finite temperature. In fact, in global thermodynamic equilibrium every normal component has to move rigidly \citep{Prix_single_vortex,Carter_Prix_Magnus}. 
Furthermore, if the heat bath does not rotate (which is our case), all the normal currents need to be at rest with respect to it, in order to minimize the free energy  at fixed winding number of  $\phi$~\cite{khalatnikov_book,Termo}. For the microscopic counterpart of this result, computed directly in a vortex configuration, see \cite{wexler1996effective}.
} and diffusive equilibrium is reached. 
This implies that the value of the circulation must be given by \eqref{thouless}, which is indeed in agreement with the result of Thouless and collaborators for the total transverse force. 

Now, let us imagine, instead, to perform the same experiment presented in Sec \ref{lointrodescribo}, but assuming that the vortex is dragged from the outer to the inner wall with a finite velocity. Since, in this case, the evolution is dissipative, equation \eqref{freeBalance} does not hold anymore. In particular, the motion of the vortex will induce deviations form the rigidity condition \eqref{comosso}, producing a non-zero correction to the circulation  $\mathcal{C}^p$, which may now deviate from the value  $\mathcal{C}^p=k Y \mathfrak{A}$ of the Thouless regime. 

If, however, the motion of the vortex is sufficiently slow (so that we are in the ``parabolic regime'' described in  \cite{gavassino2020bulk}), we can treat the deviation of the normal fluid from thermodynamic equilibrium as a slow perturbation, and it should be possible to model the dynamics of the normal component by means of the Navier-Stokes equations. 
We are, thus, facing an Oseen-type problem \cite{sonin_book_2016}, where a vortex is slowly moving in a viscous fluid. In the context of superfluid Helium this problem has been studied in detail by \citet{Thouless2001}, who found a small deviation of $\mathcal{C}^p$ from the Thouless prescription, given by an additional Iordanskii-type force $\propto (\ln u_v^y)^{-2}$.  

Finally, let us imagine that the thought experiment is performed dragging the vortex from the external to the internal wall at a speed which is high compared to the timescale at which the fluid relaxes towards thermodynamic equilibrium. In this case, the normal component will not have time to thermalise and, in the proximity of the vortex, one should rely on a kinetic description of the excitations. In this limit, the  binary and ternary collisions (in neutron stars one should also consider  collisions between particles of different species) do not have time to occur near the core of the vortex (i.e. in the ballistic region, see e.g. \cite{sonin_book_2016}). Therefore, the evolution of the  distribution function is governed by the single-quasiparticle Hamiltonian. Only under this assumption, the scattering models of \citet{sonin_book_2016} and \citet{stone2000} become valid in $^4$He and may be generalised to the neutron-star context. In this Sonin regime the proper Iordanskii force \eqref{LaIordainfdiemnticalbule} is expected to appear and we may thus impose $\mathcal{C}^p = k \chi_T/\mu_T \approx k$, as a place-holder for more refined future calculations.


We can collect together all these considerations in the foregoing expression for the Joukowski lift force (which is fully chemical-gauge invariant):
\begin{equation}\label{InFinne}
    \mathfrak{N} \mathcal{F}^J_\nu =
\begin{array}{l}
Y \mu^\rho \varpi_{\rho \nu}  \\
(Y \mu^\rho + \lambda u^\rho) \varpi_{\rho \nu} \\
b^\rho \varpi_{\rho \nu} \\
\end{array}
\quad
\begin{array}{l}
  \text{in the thermalised (Thouless) regime, } J^2 \rightarrow 0   \\
  \text{in the intermediate (Oseen) regime} \\
  \text{in the fast (Sonin) regime, } J^2  \rightarrow \infty \\
\end{array}
\end{equation}
where $\lambda \sim 1/\ln^{2}(J^2 J_2)$ and $J^2$ is the second component of $J^\nu$ in our preferred tetrad, see App \ref{tettru}. To convert the dependence on $u_v$, which is the speed at which the Wexler experiment is performed, into a dependence on $J^2$, we have used the fact that in drag models the larger $J^2$ the faster the motion of the vortices in the reference frame of $u$ is supposed to be.

We remark again that the behaviour in the Thouless limit has been derived rigorously from multifluid thermodynamics. Furthermore, the Oseen results are unlikely to be modified qualitatively by the change of physical setting. However, one still needs to verify with a microscopic model that the Sonin results extend naturally to the neutron-star context. Therefore, the third line of equation \eqref{InFinne} should be considered more an educated guess than a rigorous statement. 

Finally, note that, in this set of ideas, there seems to be no room for the widely adopted prescription of \citet{langlois98}
\begin{equation}\label{mutaformae}
       \mathfrak{N} \mathcal{F}^J_\nu = n^\rho \varpi_{\rho \nu}.
\end{equation}
However, due to the intrinsic chemical-gauge dependence of this formula, one may always fix the gauge in such a way as to mimic any of the lines of equation \eqref{InFinne}. In particular, starting from the currents $p^\nu$ and $n^\nu$, counting respectively protons and neutrons, one may perform the gauge transformation \eqref{ZZZZZZZa}. In this way
\begin{equation}
    \tilde{n}^T = n_{\text{eff}}^T,
            \label{jonasciccione}
\end{equation}
and any calculation performed using the prescription \eqref{mutaformae} remains valid. One should simply reinterpret the current $n^\nu$ as the effective neutron current
\begin{equation}
    n_{\text{eff}}^\nu = n_{\text{eff}}^T u^\nu +J^\nu,
\end{equation}
and the proton current as the rest of the baryon current. 

 Although the gauge transformation \eqref{ZZZZZZZa}  can, in principle,  be made only locally 
(it can be made globally only if $\mathcal{Z}=const$), using the model of \citet{langlois98} by interpreting the neutron density via \eqref{jonasciccione} may still constitute a good approximation for most practical purposes. 
In fact,  extensions of the hydrodynamic model discussed here to the case where the gauge-fixing parameter $a$ can vary from point to point (see e.g. \citet{carter_macro_2006}) should not lead to significant corrections in glitch models, where the currents are typically assumed to be circular \citep{sourie_glitch2017,antonelli+2018,GavassinoGeometria,montoli2020A&A}. In this particular approximation, the right hand side of equation (2.35) in \citet{carter_macro_2006} vanishes and the conservation of the ``free'' neutron current is recovered, restoring the consistency with the model of \citet{langlois98}. 



\section{Conclusions}

We have analyzed some formal aspects of the effective two-fluid hydrodynamic description of the inner crust of a neutron star initiated by \citet{langlois98}, which is based on the simplifying assumptions of absence of elastic stresses, viscosity and heat conduction. It also neglects the effects coming from the vortex-array elasticity.
Not surprisingly, as a consequence of these assumptions, it is easy to establish a connection between this relativistic two-fluid effective description and the relativistic HVBK hydrodynamics for a single superfluid at finite temperature of \citet{GusakovHVBK}. 

The new ingredient we added is the inclusion of all the possible (geometrically allowed) force terms in the phenomenological equation for vortex motion (the PEVM, i.e. equations \eqref{generaleBuf} or \eqref{lacentrale}), that allows to give a physical meaning to the purely geometric expression for the vortex velocity \eqref{lacentrale}: in fact, the two HVBK coefficients $\slashed{\alpha}$ and $\slashed{\beta}$ can be written in terms of the various coefficients that appear in the PEVM (see \eqref{alpha_beta}), which, in principle, should be more directly linked to the basic processes that give rise to dissipation and hydrodynamic lift in a vortex surroundings \cite{barenghi1983JLTP}. 

The physical interpretation of such terms is, however, not so straightforward because of a fundamental ambiguity present into the two-fluid formalism, known as chemical gauge, which has its roots into the problem of identifying  some of the confined neutrons as part of the collectively-comoving normal baryon current \cite{Carter_Starting_Point,carter_macro_2006,Termo}. 
Different choices of chemical basis (i.e. different definitions of the fundamental currents of the theory, $n^\nu$ and $p^{\nu}$) lead to different equations of motion when the action principle of \citet{Carter_Starting_Point} is applied. 
Furthermore,  different PEVM, derived by using different chemical gauge choices, always differ from one another by a Generalised Iordanskii force (i.e. the Iordanskii force rescaled by a prefactor $\mathcal{Z}$, that in principle should be computed from microphysics and is related to the circulation $\mathcal{C}^p$ of the normal component around a vortex, see \eqref{fnvjifrv}). 

Since our general PEVM contains a priori this Iordanskii force term, we analysed the controversy regarding the value of  $\mathcal{Z}$: in the context of neutron star physics this Iordanskii force term is necessary to guarantee the chemical gauge covariance of the hydrodynamic equations.

We find that, by appropriately setting  the intensity of the Generalised Iordanskii force (i.e. by imposing that $\mathcal{Z}$, or $\mathcal{C}^p$, is zero), it is possible to recover the PEVM considered by  \citet{langlois98} in their model of vortex-mediated mutual friction, see also \cite{Andersson_Mutual_Friction_2016,GavassinoGeometria}. This prescription, however, turns out to be chemical-gauge dependent and therefore ambiguous.

On the other hand, following the analysis of \citet{Carter_Prix_Magnus}, we find that it is possible to obtain two gauge-independent prescriptions for the Jukowski lift force that can be interpreted as the neutron-star analogues of respectively the \citet{sonin1997PhRvB,stone2000} result for the the Iordanskii force and the \citet{Thouless2001} model in its absence. The latter turns out to be consistent with the weak-drag limit of the HVBK equations proposed by \citet{GusakovHVBK}.

In the context of neutron stars, where the subject is even more complicated than in superfluid $^4$He, the only way to compute $\mathcal{Z}$ and solve the controversy is to make a detailed microscopic model of the vortex core, in a way to compute the circulation $\mathcal{C}^p$ of the normal momentum. 

To facilitate the inclusion of a possible Iordanskii force  in hydrodynamic models, we have shown that it is possible to encapsulate the parameter $\mathcal{Z}$ within the definition of an effective neutron density which plays exactly the same role of  the physical neutron density in the PEVM postulated by \citet{langlois98}. 
In this way, one does not need to modify the equations of e.g. a glitch model, but only to redefine the neutron density (and therefore the neutron average velocity) appropriately, similarly to what can be done with the entrainment coupling \cite{antonelli+2018,GavassinoGeometria,montoli2020A&A}. 

Finally, we revised the thought experiment of \citet{Wexler1997}, which, in combination with the early result of Thouless, allowed to conclude that there is no contribution to the Jukowski lift force (i.e. the total transverse force acting on a vortex) that is dependent on the normal fluid velocity. 
We extended the original experiment to a special relativistic context, by using the thermodynamic tools developed in \cite{Termo,GavassinoTermometri}. This leads us to  conclude that a possible resolution of the apparent mismatch between the results of Sonin and collaborators and the one of Thouless and collaborators on the Iordanskii force may be due to the fact that they are pertinent to different (and opposite) dynamical regimes (a quasi-static one, consistent with the result of Thouless, and a fast one, consistent with the result of Sonin and Stone). 

Given the still high level of uncertainty in quasiparticle and vortex kinetic theory, it is difficult to predict which regime is expected to be dominant in neutron star crusts. However, in the absence of more rigorous arguments, one may consider that the typical time-scale for the excitation of new Kelvin waves on a vortex can be estimated by considering the typical time  $\sim 10^{-15}\,$s  needed to move between two pinning sites (see Fig 5 in \citep{antonelli2020MNRAS}), which is likely to be much shorter than any possible relaxation time-scale towards thermodynamic equilibrium (although a clear estimate of such a time-scale is still unknown). 
Therefore, we are tempted to opt for the Sonin regime for neutron star applications, but defer to future work a more detailed analysis of the problem.

\vspace{6pt} 

\acknowledgments{
Partial support comes from PHAROS, COST Action CA16214. M.A. acknowledges support from the Polish National Science Centre (NCN) grant SONATA BIS 2015/18/E/ST9/00577, P.I.: B. Haskell. L.G acknowledges support from Polish National Science Centre (NCN) grant OPUS 2019/33/B/ST9/00942. This research received no external funding.
}




\appendix

\section{Recovering the standard formulation of the HVBK vortex velocity}
\label{AAA}

We show that the expression \eqref{velocitabella} for $u_v$ is equivalent to the vorticity conservation law of the HVBK hydrodynamics in the absence of vortex tension. Furthermore, we find the relationship between the coefficients $u_v^{(J)},\mathcal{D}$ and the standard HVBK kinetic coefficients $\alpha$ and $\beta$ used, for instance, by~\citet{GusakovHVBK}.
In the subsection \ref{preliminaryIdentity} we derive a preliminary geometric identity, which is then used in subsection \ref{IlcontoLab} to establish a connection with equation (57) of~\cite{GusakovHVBK}, which is the relativistic generalization of the HVBK vorticity-conservation equation.

\subsection{A preliminary identity}\label{preliminaryIdentity}

We consider the four-vector $q$, defined with the aid of the projector $h$  as
\begin{equation}
q^\nu =h^{\nu \lambda} \varpi_{\lambda \sigma} J^\sigma \,   \spc 
h^{\nu \rho} = g^{\nu \rho} + u^\nu u^\rho \, .
\end{equation}
We can use equation \eqref{explixit} to obtain the identity
\begin{equation}
q^\nu \,= \,k \,\mathfrak{N} \, h^{\nu \lambda}\,  \varepsilon_{\lambda \sigma \alpha \beta} \, u_v^\alpha \, l^\beta \, J^\sigma \, .
\end{equation}
Expanding $u_v$ according to equation \eqref{velocitabella}, we obtain
\begin{equation}
 q^\nu = k \mathfrak{N} \, h^{\nu \lambda}  \varepsilon_{\lambda \sigma \alpha \beta} l^\beta J^\sigma  \, 
 \big(\Gamma_v \, u^\alpha + u_v^{(J)} \, J^\alpha +u_v^{(l)} \, l^\alpha + \mathcal{D} \, \varepsilon^{\alpha \rho \mu \tau} J_\rho u_\mu l_\tau \big) \, . 
\end{equation} 
The last three terms in the round bracket do not give any contribution. In fact, the terms proportional to $u_v^{(J)}$ and $u_v^{(l)}$ vanish due to the antisymmetry of the Levi-Civita tensor. 
On the other hand, the triplet $l,J,\star(J \wedge u \wedge l)$ is orthogonal to $u$: when contracted with the Levi-Civita, they produce a vector which is proportional to $u$, which contracted with $h^{\nu \lambda}$ returns zero (this proves that also the term proportional to $\mathcal{D}$ vanishes). 
Hence, we find
\begin{equation}
q^\nu = k \mathfrak{N} \Gamma_v \, h^{\nu \lambda}  \varepsilon_{\lambda \sigma \alpha \beta} u^\alpha l^\beta J^\sigma 
= k \mathfrak{N} \Gamma_v \,\varepsilon\indices{^\nu _\sigma _\alpha _\beta} u^\alpha l^\beta J^\sigma
\, .  
\end{equation}
Finally, by employing equation \eqref{thermalDensityVortex}, we obtain the useful identity
\begin{equation}\label{identity}
\varepsilon^{\nu \rho \sigma  \lambda} J_\rho u_\sigma  l_\lambda  =\dfrac{1}{k\mathfrak{N}_T} h^{\nu \lambda} \varpi_{\lambda \sigma} J^\sigma \, . 
\end{equation}

\subsection{Recovering the HVBK hydrodynamics of Gusakov }
\label{IlcontoLab}

Let us define an auxiliary vector $V$ through the equation
\begin{equation}\label{auXXXXXXXXX}
u_v^\nu = \Gamma_v V^\nu +u_v^{(l)} \, l^\nu \, .
\end{equation}
If we contract this relation with $\varpi_{\nu \rho}$ and recall the formula \eqref{explixit}, we obtain
\begin{equation}\label{gabibbo}
V^\nu \varpi_{\nu \rho} =0 \, .
\end{equation}
If we manage to prove that $V$ coincides with the vector $V_{(L)}$ introduced by \citet{GusakovHVBK} in equation (57), then \eqref{gabibbo} is equation (58) of \cite{GusakovHVBK}, which he showed to be the relativistic analogue of the vorticity conservation equation in HVBK hydrodynamics (in the absence of the vortex energy-density contribution to the stress-energy tensor). 

If we combine \eqref{auXXXXXXXXX} with \eqref{velocitabella} and \eqref{identity} we immediately obtain
\begin{equation}
V^\nu = u^\nu + \dfrac{u_v^{(J)}}{\Gamma_v} J^\nu + \dfrac{\mathcal{D}}{k\mathfrak{N}_T \Gamma_v} h^{\nu \lambda} \varpi_{\lambda \sigma} J^\sigma \, . 
\end{equation}
We see that it coincides with (57) of \cite{GusakovHVBK} provided that we make the identifications
\begin{equation}\label{IiDoDi}
\dfrac{u_v^{(J)}}{\Gamma_v} = - \mu_T  \alpha  \spc   \dfrac{\mathcal{D}}{ \Gamma_v} = \mu_T \beta \, ,
\end{equation}
with $\mu_T =-\mu_\nu u^\nu$, which are the relations we were looking for.

However, we find it more convenient to work with the coefficients
\begin{equation}
\slashed{\alpha} = \mu_T  \alpha \spc  \slashed{\beta} = \mu_T  \beta \, ,
\end{equation}
which, inserted in \eqref{IiDoDi}, satisfy the relations
\begin{equation}
u_v^{(J)} = - \Gamma_v \slashed{\alpha} \spc  \mathcal{D}=\Gamma_v \slashed{\beta} \, . 
\end{equation}

\section{Tetrad calculations}

We summarise the properties of the tetrad we introduce in Sec \ref{HVBKcacloili} and perform explicitly the tetrad calculations which are omitted from the main body.

\subsection{Tetrad formulary}\label{tettru}

The tetrad $e_a$ is constructed imposing
\begin{equation}
e_0 = u  \spc e_2 = \dfrac{J -g(J,l) \, l}{\sqrt{g(J,J)-g(J,l)^2}}  \spc e_3 =l \, .
\end{equation}
The vector $e_1$ is uniquely determined by the requirements orthonormality and right-handed orientation,
\begin{equation}
e_1 = - \dfrac{\star (J \wedge u \wedge l)}{\sqrt{g(J,J)-g(J,l)^2}} \, .
\end{equation}
It is immediate to see that in this basis
\begin{equation}
\begin{split}
& u= (1,0,0,0) \\
& l= (0,0,0,1)  \\
& u_v= (\Gamma_v, \Gamma_v \Delta^1,\Gamma_v \Delta^2,0)\\
& J= (0,0,J^2,J^3) \\
& n= (n^T,0,J^2,J^3) \, . 
\end{split}
\end{equation}
We can easily write some of the components appearing in the above decomposition as covariant expressions, e.g.,
\begin{equation}
 \Gamma_v = -u_v^\nu u_\nu \spc
 J^2 =\sqrt{ J^\nu J_\nu - (J^\nu l_\nu)^2} \spc
 J^3 = J^\nu l_\nu \spc
 n^T = -n^\nu u_\nu \, . 
\end{equation}  
As a direct application, let us rewrite \eqref{lacentrale} in this tetrad; it is immediate to see that it becomes
\begin{equation}
u_v = ( \Gamma_v , \Gamma_v \slashed{\beta} J^2, -\Gamma_v \slashed{\alpha}J^2,0  ) \, .
\end{equation} 
This allows us to link the HVBK coefficients with the components of $u_v$, namely 
\begin{equation}\label{Chepassione}
\Delta^1 =  \slashed{\beta} J^2  \spc \Delta^2 = -\slashed{\alpha}J^2 \, . 
\end{equation} 

\subsection{The PEVM in the tetrad formalism}\label{tetraddizzareIldrag}

Our goal is to write explicitly the tetrad components of the general PEVM in \eqref{pufficosa}. 
Let us start with the left-hand side. As a first step, we notice that
\begin{equation}
n^T_{\text{eff}} u^a +J^a = (n^T_{\text{eff}} \, , \, 0 \, , \, J^2 \, , \, J^3) \, ,
\end{equation}
and contracting with \eqref{terbufi} we  obtain 
\begin{equation}
(n^T_{\text{eff}} u^a +J^a)\varpi_{ab} = k \mathfrak{N} \Gamma_v 
\left( 
     \Delta^1 J^2 \, , \, 
 n^T_{\text{eff}} \Delta^2 -J^2   \, , \, 
 - n^T_{\text{eff}} \Delta^1 \, , \, 
  0   \right) \, .
\end{equation}
Under the assumption of slow relative motion between the two components, we can neglect the quadratic terms in relative speed and make the approximations $\Gamma_v \approx 1$ and $\Delta^1 J^2 \approx 0$. Recalling \eqref{Chepassione}, we arrive at the final expression for the left-hand side of \eqref{pufficosa},
\begin{equation}\label{bellaleft}
(n^T_{\text{eff}} u^a +J^a)\varpi_{ab} \approx  k \mathfrak{N}  J^2
\left( 
    0 \, , \,
  -n^T_{\text{eff}} \slashed{\alpha} -1  \, , \,
  -n^T_{\text{eff}} \slashed{\beta} \, , \,
   0
  \right) \, .
\end{equation}
Now, let us move to the right-hand side of the PEVM in \eqref{pufficosa}. Considering that
\begin{equation}
\mathcal{R}' n^a + \mathcal{R} n^T_{\text{eff}} u^a = (\mathcal{R} n^T_{\text{eff}}+\mathcal{R}'n^T \, , \,0 \, , \,\mathcal{R}' J^2 \, , \,\mathcal{R}' J^3)  \, , 
\end{equation}
we immediately see (working directly in the limit of small relative speeds) that
\begin{equation}
\begin{split}
& u_{va}(\mathcal{R}' n^a + \mathcal{R} n^T_{\text{eff}} u^a)\approx -\mathcal{R} n^T_{\text{eff}}-\mathcal{R}'n^T \\
& l_a(\mathcal{R}' n^a + \mathcal{R} n^T_{\text{eff}} u^a)= \mathcal{R}' J^3. \\
\end{split}
\end{equation}
Since
$ {\perp}_{ab} = \eta_{ab} + u_{va} u_{vb} - l_a l_b  $, which is a consequence of  \eqref{paralleEel}, we finally obtain the explicit expression of the right-hand side of \eqref{pufficosa}:
\begin{equation}\label{bellaright}
k \mathfrak{N} {\perp}_{ab}(\mathcal{R}' n^a + \mathcal{R} n^T_{\text{eff}} u^a) \approx k \mathfrak{N} J^2  
\left( 
     0 \, , \,
-\mathcal{R}  n^T_{\text{eff}} \slashed{\beta} - \mathcal{R}' n^T \slashed{\beta}  \, , \,
 \mathcal{R}' + \mathcal{R}  n^T_{\text{eff}} \slashed{\alpha} + \mathcal{R}' n^T \slashed{\alpha}  \, , \,
  0 
  \right)\, .
\end{equation}
Imposing the equality of the two sides, equations \eqref{bellaleft} and \eqref{bellaright} we obtain the system \eqref{bellatutto}, which is what we wanted to prove.

\section{Mesoscopic models for the flow around a vortex}

In this appendix we discuss the mesoscopic physical assumptions that lead to respectively \eqref{hjklm}, discussed in App \ref{vbaf}, and \eqref{thouless}, discussed in App \ref{riGGG}. Here, all the quantities are considered to be mesoscopic variables, and the hydrodynamic description is applied at the inter-vortex separation scale. 
Hence, the physical setting is the one considered by \citet{Carter_Prix_Magnus}.

\subsection{Induced circulation in the normal component}\label{vbaf}

Here we discuss the implications of assuming that the motion of the vortex perturbs the normal component in such a way that
\begin{equation}\label{baruffantone}
\mu_T \oint u_\sigma dx^\sigma = k.
\end{equation}
In addition, we work under the same assumptions made by \citet{sourie2020IJMPB}, namely of homogeneous thermodynamic variables far away from the vortex.

We start with the observation that $\chi_\nu$ can be written as a linear combination of $\mu_\nu$ and $u_\nu$, since these are the only two relevant (linearly independent) covectors of the mesoscopic model,
\begin{equation}\label{cheBelzo}
    \chi_\nu = \mathcal{L} \mu_\nu + \mathcal{G} u_\nu,
\end{equation}
where $\mathcal{L}$ and $\mathcal{G}$ are two linear combination coefficients which should be treated as homogeneous constants (provided that we are sufficiently far away from the vortex core). Contracting \eqref{cheBelzo} with $-u^\nu$ we find
\begin{equation}\label{AAAAAAAABBBBBBBBBB}
    \chi_T = \mathcal{L} \mu_T + \mathcal{G}.
\end{equation}
On the other hand, if we integrate \eqref{cheBelzo} around a closed loop, far away from the vortex and surrounding the vortex only once, we obtain
\begin{equation}\label{BBBBBBBBBBBCCCCCCCCCC}
    \mathcal{C}^p = \mathcal{L}k + \mathcal{G} \dfrac{k}{\mu_T} \, ,
\end{equation}
where we have used \eqref{circolazioniIIIIIII} and \eqref{baruffantone}. Combining \eqref{BBBBBBBBBBBCCCCCCCCCC} with \eqref{AAAAAAAABBBBBBBBBB}, recalling \eqref{fnvjifrv}, we finally obtain 
\begin{equation}\label{DDDDDDCCCCC}
    \mathcal{Z} = \dfrac{\mathcal{C}^p}{k} =\dfrac{\chi_T}{\mu_T} .
\end{equation}
Given that we have used only chemical-gauge covariant assumptions, this formula must be in turn chemical-gauge covariant. This can be explicitly checked by noting that, if we contract both sides of equations \eqref{tuttiinbarca} with $\tilde{u}^\nu = u^\nu$, we obtain
\begin{equation}
\tilde{\chi}_T = (1-a)\chi_T + a \mu_T  \spc  \tilde{\mu}_T = \mu_T
\, ,
\end{equation}
whose ratio is
\begin{equation}
    \dfrac{\tilde{\chi}_T}{\tilde{\mu}_T} = (1-a) \dfrac{\chi_T}{\mu_T} + a \, .
\end{equation}
This transformation rule, however, is the same as \eqref{ZZZZZZZ}, proving the covariance of \eqref{DDDDDDCCCCC}.

\subsection{Thermodynamic equilibrium}\label{riGGG}

Here we discuss the implications of assuming that the fluid is in thermodynamic equilibrium (for a fixed position of the vortex line) in a neighbourhood of the vortex. We will show that this leads to equation \eqref{thouless} for $\mathcal{C}^p$. 

We start from the observation that in thermodynamic equilibrium the normal four-velocity is rigid. Since we know that, in the setting of \cite{Carter_Prix_Magnus}, the fluid is asymptotically non-rotating, we can conclude that there is a (macroscopically local, but mesoscopically extended) inertial frame $(t,x^1,x^2,x^3)$ surrounding the vortex line in which $u^\nu$ is static, namely
\begin{equation}
u^0 =1  \spc  u^j=0.
\end{equation}
Thus, if we pull back equations \eqref{decompbubu} on $t=\text{const}$ hypersurfaces, we obtain
\begin{equation}
\chi_j =\mathfrak{A}n_j  \spc \mu_j = Y^{-1} n_j,
\end{equation}
which implies
\begin{equation}\label{cccchhh}
\chi_j = \mathfrak{A}Y \mu_j.
\end{equation}
For the analysis of \cite{Carter_Prix_Magnus} to be valid, we need to impose (local) irrotationality of both the momenta, which is most easily realised if we require 
\begin{equation}\label{partialll}
\partial_j (\mathfrak{A}Y)=0.
\end{equation}
This condition is automatically valid in diffusive equilibrium. In fact, working in the aforementioned inertial frame, we can always write the deviation of $\mathfrak{A}Y$ from homogeneity as a perturbation, taking as independent thermodynamic variables $\chi_T$, $\mu_T$ and $w^\nu w_\nu$:
\begin{equation}
\delta (\mathfrak{A}Y) = \dfrac{\partial (\mathfrak{A}Y)}{\partial \chi_T} \delta \chi_T + \dfrac{\partial (\mathfrak{A}Y)}{\partial \mu_T} \delta \mu_T + \dfrac{\partial (\mathfrak{A}Y)}{\partial (w^\nu w_\nu)} \delta (w^\nu w_\nu) . 
\end{equation}
However, if $\chi_T$ and $\mu_T$ are homogeneous (which is the condition of diffusive equilibrium), we have to impose
\begin{equation}
\delta \chi_T =0  \spc \delta \mu_T =0,
\end{equation}
which implies
\begin{equation}
\delta (\mathfrak{A}Y)  = \dfrac{\partial (\mathfrak{A}Y)}{\partial (w^\nu w_\nu)} \delta (w^\nu w_\nu).
\end{equation}
Since this is a second order in the relative velocity between the species we can neglect it, proving that $\mathfrak{A}Y$ is homogeneous up to the first order.

Finally, taking a loop (surrounding the vortex once) which lies entirely inside $t=\text{const}$ hypersurface, we can combine \eqref{cccchhh} and \eqref{partialll} to obtain
\begin{equation}
\mathcal{C}^p = \mathfrak{A}Y k,
\end{equation}
which is what we wanted to prove.
\reftitle{References}


\externalbibliography{yes}
\bibliography{biblio}





\end{document}